\documentclass[sigconf, nonacm]{style/acmart}

\usepackage{microtype}
\usepackage{xstring}

\usepackage{pgfplots}
\pgfplotsset{compat=1.18}
\usepgfplotslibrary{fillbetween}
\usetikzlibrary{intersections}
\usetikzlibrary{positioning}
\usepgfplotslibrary{statistics}
\usetikzlibrary{matrix}
\usepgfplotslibrary{groupplots}

\usepackage{amsmath}
\usepackage{siunitx}
\usepackage{bbm}

\usepackage[scaled=0.825]{beramono}
\usepackage[T1]{fontenc}

\usepackage{xcolor}
\definecolor{codeColor}{RGB}{20, 110, 110}
\definecolor{codeDarkColor}{RGB}{0, 60, 60}
\definecolor{linkColor}{RGB}{20, 20, 100}
\definecolor{citeColor}{RGB}{100, 20, 20}

\usepackage[]{hyperref}
\hypersetup{
  colorlinks,
  linkcolor={red!40!black},
  citecolor={red!40!black},
  urlcolor={red!40!black}
}
\usepackage[nameinlink,capitalise,noabbrev]{cleveref}
\crefname{lstlisting}{listing}{listings}
\Crefname{lstlisting}{Listing}{Listings}

\usepackage{listings}

\usepackage{subcaption}
\usepackage{caption}

\usepackage{adjustbox}

\usepackage{tabularx, booktabs}
\usepackage{multirow}

\usepackage{ifthen}

\usepackage{stfloats}

\usepackage{array}

\lstdefinestyle{sexp}{
    language=Lisp,
    showstringspaces=false,
    basicstyle=\color{codeDarkColor}\ttfamily,
    keywordstyle=\color{codeColor}\ttfamily,
    stringstyle=\color{red}\ttfamily,
    identifierstyle=\color{codeColor},
    mathescape=true
}

\definecolor{poscolor}{RGB}{128, 180, 230}
\definecolor{negcolor}{RGB}{252, 90, 40}
\definecolor{neucolor}{RGB}{1, 74, 26}
\definecolor{cmarkcolor}{RGB}{11, 49, 66}
\definecolor{xmarkcolor}{RGB}{170,72,72}

\definecolor{colunixcoder}{RGB}{252, 90, 40}
\definecolor{colpdbert}{RGB}{128, 180, 230}
\definecolor{colcodegen}{RGB}{1, 74, 26}
\definecolor{collinevul}{RGB}{152, 190, 87}
\definecolor{colstarcoder}{RGB}{11, 49, 66}

\usepackage{pifont} 
\usepackage{wasysym} 
\usepackage{xspace}

\usepackage{enumitem}
\setlist[itemize]{itemsep=4pt}

\newcommand{\tikzcircle}[2][red,fill=red]{\tikz[baseline=-0.5ex]\draw[#1,radius=#2] (0,0) circle ;}
\newcommand{\cbstart}{\color{black}}
\newcommand{\cbend}{\color{black}}
\newcommand{\cb}[1]{\textcolor{black}{#1}}
\newcolumntype{R}[1]{>{\raggedleft\arraybackslash}p{#1}}
\newcommand{\positive}[1]{\textcolor{cmarkcolor}{#1}}
\newcommand{\negative}[1]{\textcolor{xmarkcolor}{#1}}
\newcommand\blfootnote[1]{%
  \begingroup
  \renewcommand\thefootnote{}\footnote{#1}%
  \addtocounter{footnote}{-1}%
  \endgroup
}

\newcommand{\metric}{\mu}
\newcommand{\metrics}{\mathcal{M}}

\newcommand{\score}{s}
\newcommand{\code}{c}
\newcommand{\codes}{C}
\newcommand{\tree}{t}
\newcommand{\trees}{T}
\newcommand{\filter}{F}
\newcommand{\map}{M}

\newcommand{\reduce}{R}

\newcommand{\subtree}{t}

\newcommand{\scmAuto}{SCM}

\ignorespacesafterend
\newcommand{\mi}{\text{I}}

\lstnewenvironment{sexp}{\lstset{
    style=sexp,
    aboveskip=-7pt,
    belowskip=-7pt,
    framextopmargin=0pt,
    frame=none
}}{}
\newcommand{\sexpinline}[1]{\lstinline[style=sexp]/#1/}

\usepackage{pifont} 
\newcommand{\cmark}{\positive{\small\ding{51}}}
\newcommand{\xmark}{\negative{\small\ding{55}}}

\newcommand{\best}[1]{\underline{\textbf{#1}}}
\newcommand{\secondbest}[1]{\textbf{#1}}


\thanks{\textcopyright \hspace{1pt}Felix Weissberg, Lukas Pirch, Erik Imgrund, Jonas Möller, Thorsten Eisenhofer, and Konrad Rieck, 2025. This is the author's version of the work. It is posted here for your personal use. Not for redistribution. The definitive Version of Record was published in the proceedings of the 48th ACM/IEEE International Conference on Software Engineering (ICSE '26), \href{https://doi.org/10.1145/3744916.3764574}{https://doi.org/10.1145/3744916.3764574}.}

\begin{document}
\date{}

\title[LLM-based Vulnerability Discovery through the Lens of Code Metrics]{LLM-based Vulnerability Discovery through\\ the Lens of Code Metrics}

\author{\vspace{0.1cm}Felix Weissberg*}
\affiliation{%
  \institution{BIFOLD \& TU Berlin}
  \city{Berlin}
  \country{Germany}
}

\author{\vspace{0.1cm}Lukas Pirch*}
\affiliation{%
  \institution{BIFOLD \& TU Berlin}
  \city{Berlin}
  \country{Germany}
}

\author{\vspace{0.1cm}Erik Imgrund}
\affiliation{%
  \institution{BIFOLD \& TU Berlin}
  \city{Berlin}
  \country{Germany}
}

\author{\vspace{0.1cm}Jonas Möller}
\affiliation{%
  \institution{BIFOLD \& TU Berlin}
  \city{Berlin}
  \country{Germany}
}

\author{\vspace{0.1cm}Thorsten Eisenhofer}
\affiliation{%
  \institution{BIFOLD \& TU Berlin}
  \city{Berlin}
  \country{Germany}
}

\author{\vspace{0.1cm}Konrad Rieck}
\affiliation{%
  \institution{BIFOLD \& TU Berlin}
  \city{Berlin}
  \country{Germany\vspace{0.1cm}}
}

\renewcommand{\shortauthors}{
    Felix Weissberg,
    Lukas Pirch,
    Erik Imgrund,
    Jonas Möller,
    Thorsten Eisenhofer,
    and Konrad Rieck
}

\renewcommand{\authors}{Felix Weissberg,
Lukas Pirch,
Erik Imgrund,
Jonas Möller,
Thorsten Eisenhofer,
and Konrad Rieck
}

\begin{abstract}
Large language models (LLMs) excel in many tasks of software engineering, yet progress in leveraging them for vulnerability discovery has stalled in recent years. To understand this phenomenon, we investigate LLMs through the lens of classic code metrics. Surprisingly, we find that a classifier trained solely on these metrics performs on par with state-of-the-art LLMs for vulnerability discovery. A root-cause analysis reveals a strong correlation and a causal effect between LLMs and code metrics: When the value of a metric is changed, LLM predictions tend to shift by a corresponding magnitude. This dependency suggests that LLMs operate at a similarly shallow level as code metrics, limiting their ability to grasp complex patterns and fully realize their potential in vulnerability discovery. Based on these findings, we derive recommendations on how research should more effectively address this challenge.
\end{abstract}

\maketitle

\section{Introduction}
\label{sec:introduction}
Machine learning models are on the verge of becoming an integral part of software development, supporting essential tasks like code completion, refactoring, and auditing. A particularly crucial aspect in this context is the identification of security vulnerabilities \cb{early during the development process}. Over the past decade, several methods have been proposed to detect security flaws using machine learning, ranging from simple classifiers~\mbox{\citep{GruWasZel10, YamLotRie12, ScaWlHovJoo14,PerDecSmiArp+15}} to large language models (LLMs) for code~\citep{FenGuoTanDua+20, GuoLuDuaWan+22, LoLi+24, WanLeGotBui+23}. However, despite increasingly larger models and broader training data, progress in vulnerability discovery has recently begun to stall, indicating challenges in unlocking the \cb{full} potential of machine learning for this task~\citep{CheDinAloChe+23, DinFuIbrSit+25}.

In this paper, we take a step back and ask: After a decade of research, what progress have we made compared to minimal baselines?
To this end, we contrast LLMs with a classic tool of software engineering: \emph{code metrics}. Originating from the early days of programming, these metrics quantify basic properties of software, such as the lines of code, the nesting level of loops, or the number of goto statements~\citep{Hal77,McC76,HarMagKluc+82,ShiWil08}.
In particular, we focus on a generalized family of \emph{syntactic code metrics}, encompassing both historical and current metrics derived from syntax alone~\mbox{\citep{MenGueMacAgh+21,LiuMenZouGon+20,ShiWil+13,DuCheLiGuo+19}}. 
Although these metrics lack the sophistication of LLMs, their simplicity offers an interesting lens for evaluating progress---a viewpoint that has not been explored so far.

For this retrospective investigation\blfootnote{*Authors contributed equally.}, we consider five state-of-the-art LLMs for vulnerability discovery and generative code tasks: PDBERT~\citep{LiuTanZhaXia+24}, UniXcoder~\citep{GuoLuDuaWan+22}, LineVul~\citep{FuCha+22}, CodeGen~2.5~\citep{NiHaXiSa+23}, and StarCoder~2~\citep{LoLi+24}. Among these, UniXcoder is currently regarded as the most effective approach for learning-based vulnerability discovery, following a comprehensive evaluation of LLMs \cb{for this task}~\citep{DinFuIbrSit+25}. We compare these models against \cb{a set of 23 syntactic} code metrics through a series of experiments on PrimeVul~\citep{DinFuIbrSit+25}, the latest benchmark dataset for vulnerability discovery \cb{approaches}, consisting of 220,000 labeled C/C++ functions from \cb{more than 750} open-source projects.

Our findings challenge the prevailing paradigm in prior work that large and sophisticated learning models excel over simple baselines in vulnerability discovery. \cb{On the contrary, we find that} a classifier trained solely on syntactic code metrics performs on par with the best current LLMs. In other words, state-of-the-art performance can be achieved using simple statistical features of code and just 6\% of the model parameters, as shown in \cref{tab:intro-overview}.
Surprisingly, even a classifier trained \cb{only} on a single metric achieves almost the same effectiveness, reaching \cb{over 90\%} of the LLMs' performance. 
Neither the complex mechanisms of the transformer architecture nor the vast number of model parameters can be fully exploited to significantly outperform the performance of this simple metric, indicating a notable discrepancy to prior work on \cb{source-code} vulnerability discovery.

To gain insights into these unexpected results, we conduct a root-cause analysis and examine causal dependencies between LLMs and code metrics within the framework of \citet{Pea09}. First, we demonstrate that LLMs can reproduce code metrics and hence have access to the same underlying information. Second, we show that combining LLMs with code metrics does not lead to any improvement, suggesting that they rely on overlapping rather than complementary information. Finally, we demonstrate that the predictions of the best LLMs are not only strongly correlated with code metrics but also causally dependent: When the value of a metric is changed, the LLM predictions tend to shift by a proportional magnitude.	This dependency indicates that both approaches rely on equivalent---though not necessarily identical---information for their decision-making process.
    
While we cannot fully open the black box of LLMs, our analysis reveals that these models do not access information fundamentally different from counting lines or loops in code. The models operate at a surface level of the code, focusing on basic statistical properties rather than uncovering genuine patterns of vulnerabilities. This outcome is both unexpected and disappointing. It raises questions about the considerable resources required for training and deploying LLMs and underscores the need for a deeper understanding of their capabilities in code analysis.

Based on this new perspective, we derive recommendations for the research community: First, code metrics need to be adopted as simple baselines, serving as a sanity check for analyzing performance improvements. To foster this development, we make our implementation and experimental framework publicly available\footnote{\url{https://github.com/mlsec-group/cheetah}}. Second, we should strive for a more balanced perspective when evaluating learning models for vulnerability discovery, ensuring that model complexity is always assessed in relation to detection results. Finally, we must aim to develop learning models that move beyond surface-level statistics, likely necessitating a rethinking of the entire discovery process.

\medskip
\noindent In summary, we make the following contributions:
\begin{itemize}
    \setlength{\itemsep}{2pt}

    \item \textit{Retrospective evaluation.} We systematically evaluate state-of-the-art LLMs for vulnerability discovery against code metrics, uncovering a surprising similarity in their detection performance
    ($\rightarrow$ \cref{sec:evaluation}).

    \item \textit{Unified code metrics.} As basis for our investigation, we introduce a generalized family of code metrics for vulnerability discovery, unifying past and recent work on measuring properties of syntax ($\rightarrow$~\cref{sec:codemetrics}).

    \item \textit{Root-cause analysis.} We analyze the decision making of recent language models through the lens of code metrics and derive recommendations for improving current research
    ($\rightarrow$~\cref{sec:interpretability,sec:discussion}).

\end{itemize}

Note that our work is not intended as a critique of prior research efforts. Rather, we reveal that current approaches have deviated from their intended goals, hindering progress in detecting vulnerable code using machine learning. With our findings and recommendations, we are optimistic that this new perspective will help to better gauge progress and inspire more effective methods for vulnerability discovery.

\renewcommand{\arraystretch}{0.8}
\begin{table}[t]
    \caption{Performance of different vulnerability discovery approaches on the PrimeVul dataset. Factors are relative to UniXcoder, the best performing model in our experiments.}
    \label{tab:intro-overview}
    \centering
    \begin{tabular}{l rrrr}
    \toprule
    \textbf{Model} & \multicolumn{1}{c}{\textbf{F1 score $\uparrow$}} & \hspace{-2mm}\textbf{Factor} & \multicolumn{1}{c}{\textbf{\#param $\downarrow$}} & \hspace{-2mm}\textbf{Factor} \\
    \midrule
    UniXcoder~\cite{GuoLuDuaWan+22} & 20.69 $\pm$ 1.43  & 1.00 & 125M & 1.00 \\
    CodeGen 2.5~\cite{NiHaXiSa+23} & 18.57 $\pm$ 0.54 & 0.90 & 7B & 53 \\
    GPT-4o & 5.31 $\pm$ 0.35 & 0.26 & $ \approx$ 1T & 8000 \\
    \midrule
    Code metrics & 20.32 $\pm$ 0.59 & 0.98 & 7M & 0.06\\
    \bottomrule
\end{tabular}

\end{table}

\section{LLM-based Vulnerability Discovery}
\label{sec:background}
The discovery of vulnerabilities through static program analysis is a long-standing and notoriously difficult problem in software engineering. Since a universal detection approach is generally unattainable~\citep{Ric53}, research initially focused on specific defect types, such as buffer overflows~\citep{DorRodSag03, LarEva01, WagFosBreAik+00}, integer issues~\mbox{\citep{DieLiRegAdv+12, CokHaf13, WreYamMaiRie+16}}, or taint-style vulnerabilities~\citep{YamGolArpRie+14, BacRieSkoSto+17}. Recent advances in machine learning, however, have sparked optimism that broader detection methods can be developed by automatically inferring patterns of insecure code from past vulnerabilities. This optimism has been further fueled by the impressive capabilities of LLMs, which have excelled across various
application
domains. In the following, we briefly review this line of research, discussing relevant approaches and
benchmark
datasets.

\subsection{Discovery Approaches}
\label{sec:approaches}

Learning to detect vulnerabilities is a challenging problem that has evolved significantly over time. Early approaches have relied on classic concepts from text mining and information retrieval to learn from vulnerabilities. For instance, \citet{GruWasZel10}, \citet{YamLotRie12}, and \citet{ScaWlHovJoo14} use simple bag-of-words and abstract syntax tree representations to identify insecure or anomalous code. However, due to the complexity of many vulnerability types, these initial methods lack the conceptual depth required for broader detection. Hence, over the past decade, numerous studies have addressed these limitations, advancing both data representations and learning models.

\paragraph{Graph-based models} One branch of this research has specifically focused on enhancing data representations by working with code graphs~\citep[e.g.,][]{ZhoLiuSioDu+19, ChaKriDinRay+22, GanImgHärRie+23, CheWanHuaXu+21}. These graphs capture the syntax, control flow, and data flow of code in a unified representation, providing a rich basis for inferring discriminative patterns. Two prominent examples of this approach are \emph{Devign}~\citep{ZhoLiuSioDu+19} and \emph{ReVeal}~\citep{ChaKriDinRay+22}, both of which use gated graph neural networks to embed code graphs and identify vulnerabilities
on a function level.
Although these models showed state-of-the-art performance at the time when they were introduced, more recent work shifted towards language models for this task.
Compared to GNN-based approaches, they require significantly less pre-processing of the data, since \cb{accurate parsing and} control or data flow analyses are generally not required. At the same time, they have been shown to outperform the detection performance of graph neural networks~\citep{CheDinAloChe+23}.

\paragraph{Large language models} More recent methods for vulnerability discovery therefore leverage pre-trained language models for code and augment them with a classification head~\citep{FenGuoTanDua+20, GuoLuDuaWan+22, WanWanJotHoi21, WanLeGotBui+23, LoLi+24, NiHaXiSa+23}.
In this work, we consider five state-of-the-art models of this research branch:
\emph{LineVul}~\citep{FuCha+22}, \emph{PDBERT}~\citep{LiuTanZhaXia+24}, \emph{UniXcoder}~\citep{GuoLuDuaWan+22}, \emph{CodeGen~2.5}~\citep{NiHaXiSa+23}, and \emph{StarCoder~2}~\citep{LoLi+24}.
The first three are based on the RoBERTa architecture, which has shown the best performance for vulnerability discovery in a recent comparison of language models~\citep{DinFuIbrSit+25}. \cb{We use all three of them as encoder-only models.} The last two follow the trend of resorting to larger models and a transformer decoder-only architecture. Besides their architecture and size, the models differ in terms of their training tasks. For example, LineVul uses masked language modeling and denoising tasks for both code and natural language inputs and UniXcoder augments this with code completion as an additional training objective. StarCoder~2 and CodeGen~2.5 take this a step further relying solely on code completion for pre-training.

An alternative to fine-tuning specialized language models is the use of large, general-purpose models. For example, \citet{DinFuIbrSit+25} leverage models from the GPT family, such as GPT-4 from OpenAI, combined with chain-of-thought prompting to detect vulnerable code samples. We consider these general approaches as an additional baseline in our comparative evaluation (\cref{sec:comparative-analysis}).

\subsection{Benchmark Datasets}
\label{sec:datasets}

As with any application of machine learning, evaluating its efficacy requires representative data of high quality and quantity. Consequently, datasets for vulnerability discovery have been actively developed and improved in recent years. While initial evaluations relied on synthetic datasets like SARD~\citep{sard} and code labeled by static analyzers~\citep{RusKimHamLaz+18}, recent work has shifted toward constructing benchmark data from publicly available sources of real-world code, such as vulnerability reports and security patches~\mbox{\citep{ZhoLiuSioDu+19, ChaKriDinRay+22, FanLiWanNgu20, NikDriLouMit21, BhaNasMoo21, WeiMölGanImg+24}}. The resulting datasets comprise thousands of functions from open-source projects both before and after security patches were applied, effectively capturing the differences between vulnerable and non-vulnerable code samples.

\paragraph{PrimeVul.} In this work, we employ the most recent benchmark dataset for vulnerability discovery \emph{PrimeVul}~\citep{DinFuIbrSit+25} (ICSE 2025). It combines several previous datasets, including \emph{BigVul}~\citep{FanLiWanNgu20}, \emph{CrossVul}~\citep{NikDriLouMit21}, \emph{CVEfixes}~\citep{BhaNasMoo21} and the largest available dataset \emph{DiverseVul}~\citep{CheDinAloChe+23}. 	These datasets have been merged and refined through deduplication and commit-filtering techniques to ensure high-quality labeling of secure and vulnerable code. The resulting dataset comprises 224,533 functions from 755 open-source projects, where 6,062 functions contain vulnerabilities. Unlike previous datasets, it enables a more realistic estimation of classifier performance by using a temporal split of commits between training and testing.

\section{Metrics for Code}
\label{sec:codemetrics}
As a counterpoint to LLMs for vulnerability discovery, we consider a classic tool of software engineering: \emph{code metrics}. Originating over four decades ago~\citep{Hal77, McC76}, these metrics offer quantitative insights into various aspects of software design, implementation, and deployment. For example, they can encompass dynamic measures, such as execution time and test suite coverage, as well as static measures, like complexity or coupling and cohesion between components. Moreover, these quantities can be measured at different levels of granularity, ranging from individual functions in a program to entire software modules.

For our investigation, we concentrate on code metrics that analyze the syntactic structure at the function level, consistent with common approaches for learning-based vulnerability discovery. Numerous metrics fall into this category, including historic complexity measures~\citep{Hal77, McC76}, as well as recent approaches designed to indicate insecure code~\citep{DuCheLiGuo+19, MenGueMacAgh+21, LiuMenZouGon+20, ShiWil+13}. Interestingly, although these metrics capture a broad spectrum of properties, they differ only in the syntactic elements they analyze and how they aggregate the collected information. Based on this observation, we introduce a generalized family of \emph{syntactic code metrics (SCM)}, which serves as the primary tool for our study.

\subsection{A Family of Metrics}

To unify the calculation of different code metrics, we propose to express them using the \emph{filter-map-reduce} paradigm from functional programming. Given a syntax tree, a syntactic code metric first \emph{filters} relevant subtrees, \emph{maps} them to numerical values, and then \emph{reduces} the results into a single quantity. Note that specific metrics may require either abstract or concrete syntax trees for their calculation, both of which are supported by our framework and are therefore not explicitly differentiated in the following.

Formally, this calculation can be defined as a function $\metric$ that assigns a numerical score $\score \in \mathbb{R}$ to a piece of code $\code \in \codes$, based on its syntax tree $\tree \in \trees$ as follows
\begin{align*}
    \metric: \trees \mapsto \mathbb{R}, \quad \metric (\tree) = (\reduce \circ \map \circ \filter)(\tree), 
\end{align*}
where $\filter$ represents a filter function over subtrees, $\map$ a map function, and $\reduce$ the final reduction to the code metric.

\paragraph{Filter function.} 
The function $\filter$ traverses a syntax tree and returns all subtrees that satisfy a specified predicate. For example, the function may return specific statements, function calls, or control structures. To provide a unified interface for filtering, we introduce a query language to define predicates, similar to how regular expressions identify patterns in text. Specifically, we define a variant of \emph{S-expressions} for querying syntax trees in our framework. A detailed description of this process and the underlying expressions is provided in Appendix~\ref{sec:appendix-metrics}.

\paragraph{Map function.} After filtering, each selected subtree is assigned a value using the function $\map$. This mapping produces a numerical quantity that reflects a specific property of each subtree, such as its presence, depth, size, or complexity. For example, when counting the number of goto statements in a piece of code, the filter $\filter$ first collects all instances of these statements, while the map $\map$ simply returns 1 for each occurrence.

\paragraph{Reduce function.} Finally, the values returned from the mapping step are combined into a single score using the function $\reduce$. This reduction iteratively applies a binary operator to aggregate the values into the final metric output. Common reduction functions include the maximum, sum, and average of the mapped values. In our example of counting goto statements, the reduction $\reduce$ simply corresponds to the sum.

\bigskip

Although the filtering step is essential for isolating specific code constructs, the importance of the map and reduction steps becomes evident when computing more complex aggregates. Several advanced code metrics, such as cyclomatic complexity, maximum loop nesting depth, the number of heap allocations, or the number of pointer dereferences, can be expressed within our framework through a combination of filters, maps, and reductions, as shown in Appendix~\ref{sec:appendix-metrics}.

\subsection{Implementation of Metrics}
We continue to present the specific code metrics implemented in our framework. Our goal is to capture a diverse range of characteristics that are relevant to understanding and identifying software vulnerabilities. To achieve this, we analyze, unify, and extend several code metrics proposed in prior research~\cite{DuCheLiGuo+19, MenGueMacAgh+21, LiuMenZouGon+20, ShiWil+13}. Through this process, we define 23 distinct syntactic code metrics, organized into four categories.

\paragraph{S: Code smell metrics (5).} The first category includes code metrics for code smells---patterns in code that  indicate underlying design or implementation issues. In particular, we implement metrics for the number of magic numbers (S1), goto~statements (S2), and function pointers (S3). Additionally, we include a metric for function calls with unused return values (S4), which may indicate overlooked errors, as well as a metric counting if-statements without corresponding else branches (S5), which may be indicative of incomplete program logic.

\paragraph{C: Complexity metrics (12).} Second, we consider metrics that directly measure code complexity~\cite{DuCheLiGuo+19}. Specifically, we build implementations for the cyclomatic complexity (C1), which quantifies the number of linearly independent control flow paths, as well as metrics for the number and nesting level of loops (C2–C4). Furthermore, we utilize metrics for the number of function parameters (C5), the complexity of control structures (C6–C8), the number of return statements (C9), type casts (C10), local variables (C11), and the maximum number of operands in an expression (C12).

\paragraph{M: Memory metrics (3).} Third, we design code metrics that measure the frequency and nature of memory operations. Specifically, we implement metrics for the number of heap allocations (M1), pointer dereferences (M2), and pointer arithmetic operations (M3). These metrics are particularly relevant to vulnerability discovery, as memory-related operations are a common source of security flaws. For example, frequent heap allocations may suggest potential memory management issues, such as leaks or improper deallocations.

\paragraph{T: Syntax tree metrics (3).} Finally, we consider metrics characterizing the syntax tree itself, providing structural insights into the underlying code. Specifically, we implement metrics for the number of tree nodes (T1), the maximum height of the tree (T2), and the average number of children per node (T3).

\medskip
In the appendix, we provide a detailed explanation of each metric, including how they can be represented within our filter-map-reduce framework and the query language based on S-expressions.

\subsection{Learning with Code Metrics}

While individual code metrics provide valuable insights into a given piece of code, integrating them within a classifier allows us to consider various combinations and weightings, further enhancing our analysis. To this end, we compile the 23 selected metrics into a numerical feature vector,
$$
(\mu_1, \ldots, \mu_{23}) \in \mathbb{R}^{23}
$$
For each function in the PrimeVul dataset, we then construct a labeled feature vector, providing the necessary setup for training and evaluating a supervised learning model.

In contrast to the token sequences used in LLMs, this feature representation is straightforward to process with most learning algorithms, accommodating both traditional and modern classifiers. We use the Auto-ML framework~\citep{FeuKleEggSpr+15, FeuEggFalLin+22}, which automatically constructs an effective classifier for tabular data by optimizing both the choice of learning models and their hyperparameters. Furthermore, the framework automatically applies data scaling and constructs ensembles of models where necessary.

Specifically, the classifier is constructed through an automated search over a set of fifteen traditional learning models, including fully connected neural networks, random forests, support vector machines, logistic regression, and others. The resulting model uses a mixed ensemble of these models and reaches a total of 7~million parameters when all components are aggregated. Although this model is relatively large considering the limited input space, it remains minuscule compared to the LLMs in our evaluation, which contain up to billions ($10^9$) or even trillions ($10^{12}$) of parameters.

\section{Retrospective Evaluation}
\label{sec:evaluation}
\begin{table}[t]
    \centering\small
    \caption{Performance of different vulnerability prediction approaches on the PrimeVul~\citep{DinFuIbrSit+25} dataset.}
    \setlength{\tabcolsep}{6pt}
\begin{tabular}{l r r r }
    \toprule
    \textbf{Predictor}& \textbf{Parameters} & \textbf{\hspace{1em}F1 $\uparrow$} & \textbf{AUPRC $\uparrow$} \\
    \midrule

    \multicolumn{4}{l}{\emph{Code Metrics}} \\
     \phantom{xx}\scmAuto{} & $6.78 \times 10^6$ & \secondbest{20.32} $\pm$ 0.59 & \best{13.80} $\pm$ 0.82 \\

    \midrule
    
    \multicolumn{4}{l}{\emph{Code Language Models}} \\
    \rule{0pt}{2ex}
    \phantom{xx}UniXcoder~\cite{GuoLuDuaWan+22} & $1.25 \times 10^8$ &  \best{20.69} $\pm$ 1.43 & 13.32 $\pm$ 1.22 \\
    \phantom{xx}PDBERT~\cite{LiuTanZhaXia+24} & $1.25 \times 10^8$ & 19.50 $\pm$ 1.17 & \secondbest{13.38} $\pm$ 0.68 \\
    \phantom{xx}CodeGen 2.5~\cite{NiHaXiSa+23} & $6.69 \times 10^9$ & 18.57 $\pm$ 0.54 & 13.22 $\pm$ 0.30 \\
    \phantom{xx}LineVul~\cite{FuCha+22} & $1.25 \times 10^8$ & 18.48 $\pm$ 1.23 & 11.68 $\pm$ 0.82 \\
    \phantom{xx}StarCoder 2~\cite{LoLi+24} & $7.17 \times 10^9$ & 17.09 $\pm$ 0.41 & 12.46 $\pm$ 0.40 \\
    
    \midrule
    
    \multicolumn{4}{l}{\emph{General-purpose LMs}} \\ 
    \phantom{xx}GPT-3.5 Turbo & $ \approx 10^{10}$ & 5.80 $\pm$ 0.83 & 2.56 $\pm$ 0.18 \\
    \phantom{xx}GPT-4o & $ \approx 10^{12}$ & 5.31 $\pm$ 0.35 & 2.50 $\pm$ 0.10 \\

    \phantom{xx}Random & \multicolumn{1}{c}{-} & 4.42 $\pm$ 0.19 & 2.33 $\pm$ 0.01 \\

    \bottomrule
\end{tabular}
\setlength{\tabcolsep}{6pt}
    \label{tab:class_results}
\end{table}

Equipped with a unified representation of code metrics and a resulting classifier, we are ready to put LLMs to the test by comparing their performance to this simple baseline. Unlike typical evaluations that measure relative differences between recent approaches, this retrospective view enables us to assess progress through the lens of long-standing code features and identify where and how improvements have occurred over the last decades.

\subsection{Experimental Setup}

As a basis for this evaluation, we fine-tune PDBERT, LineVul, CodeGen~2.5, StarCoder~2, and UniXcoder for 10 epochs and select the best performing models based on the validation loss. The code metric model is trained with a 10-minute time budget, and its best candidate is selected based on validation performance as well.
For the general-purpose language models, we consider GPT-4o and GPT-3.5~Turbo and reproduce the chain-of-thought experiment for vulnerability classification from \citet{DinFuIbrSit+25}. To limit the costs associated with this experiment, we conduct it on a stratified random subset of 2000 samples from PrimeVul's test split.

To ensure robustness of the evaluation and account for randomness, we repeat this process 10 times for each model, employing non-exhaustive cross-validation.
The training of all models is performed on a computing cluster using 20 GPUs (NVIDIA A100 class) and 700 CPU cores. Each training run utilizes one GPU and four CPU cores for the LLM-based approaches, while the Auto-ML models are trained using four CPU cores without GPU acceleration.

\subsection{Comparative Analysis}
\label{sec:comparative-analysis}

The results of our comparative analysis on PrimeVul are presented in \cref{tab:class_results}. We report the F1 and AUPRC (area under precision-recall curve) scores, along with the corresponding $90\%$ confidence interval. We select these measures over alternatives, such as true-positive rate and false-positive rate, because they are better suited for analyzing imbalanced datasets~\citep{ArpQuiPenWar+22}, as is the case in our study. Moreover, we provide the number of parameters as a reference for the complexity of the learning models.

As a first observation, we find that all models face significant challenges in effectively solving the task of vulnerability discovery. Even the best-performing models achieve F1 scores of only around $20\%$, highlighting substantial room for improvement. However, when compared to the performance of random guessing with an F1 score of about 4\%, the models still show a clear improvement, indicating that several vulnerabilities can be identified successfully.

Upon closer examination, we find that LLMs do not demonstrate statistically significant improvements over our simple classifier based on code metrics. In the experiment, the fine-tuned UniXcoder model achieves the highest F1 score of $20.69\%$, but its mean performance is surprisingly close to that of the code metrics classifier with an F1 score of $20.32\%$. LineVul, CodeGen~2.5, StarCoder~2, and PDBERT fall short of this performance. With respect to AUPRC, the SCM-based classifier as well as PDBERT, UniXcoder and CodeGen~2.5 perform almost identically with a maximum difference in mean performance of $0.58$ between the first and fourth best model.
Considering the $90\%$ confidence interval, the results of the best approaches are hardly distinguishable. 

This result is counterintuitive, considering the significantly larger parameter sizes of state-of-the-art models compared to the classifier based on code metrics.
Consistent with this finding, we also observe that general-purpose language models are outperformed by all other approaches. Specifically, GPT-4o and GPT-3.5 Turbo exhibit the weakest performance. While this may also seem unexpected given their sheer size, these models were not tuned for this task and therefore cannot compete with specialized LLMs.

\medskip

Our analysis uncovers an unexpected phenomenon: Despite their fundamentally different data representations and model architectures, vulnerability discovery approaches based on language models and code metrics provide comparable performance on a state-of-the-art benchmark for vulnerability discovery. In other words, 23~input dimensions and an ensemble of basic classifiers give rise to a viable competitor to a pre-trained transformer model with 125 million parameters and a vocabulary of 60 thousand tokens.

\subsection{Analysis of Code Metrics}
Our findings challenge a prevailing theme in prior research, which focuses on increasing model size and complexity. To better understand these results, we conduct a detailed analysis of the predictive performance of the syntactic code metrics.

\begin{figure}[b]
    \centering
    \scalebox{0.85}{\begin{tikzpicture}
\pgfplotsset{every tick label/.append style={font=\small}}
\begin{axis}[
    xlabel={Metric},
    ylabel={\% \,of \,F1},
    axis x line*=bottom,
    axis y line*=left,
    xmin=-0.5,
    xmax=22.5,
    ymax=100.5,
    xtick={0,1,2,3,4,5,6,7,8,9,10,11,12,13,14,15,16,17,18,19,20,21,22},
    ytick={0,25,50,75,100},
    xticklabels={C11,T1,C6,S5,C1,M2,M3,C8,C2,S4,C3,C4,C7,T2,S1,S2,C10,C12,C9,M1,T3,C5,S3},
    xticklabel style={rotate=90},
    width=1.13\columnwidth,
    height=0.62\columnwidth,
    ymajorgrids,
    legend cell align={left},
    legend style={at=({0.03,0.075}),anchor=south west},
]
\addplot [mark=x, negcolor, only marks] table {data/metric_analysis/best_metrics_individually.dat};
\addlegendentry{\small{\ Isolation}}

\addplot [mark=x, poscolor, only marks] table {data/metric_analysis/best_metrics_leftout.dat};
\addlegendentry{\small{\ Leave-one-out}}
\end{axis}
\end{tikzpicture}}
    \vspace{-0.175cm}
    \caption{Analysis of the performance impact when trained solely on one code metric (red \raisebox{0.1ex}{\tikzcircle[negcolor!60, fill=negcolor!60]{2pt}}), and when trained on all but one code metric (blue \raisebox{0.1ex}{\tikzcircle[poscolor!60, fill=poscolor!60]{2pt}}).}
    \Description{A scatter plot showing the classification performance impact of individual code metrics. The y-axis represents the percentage of F1 score, ranging from 0 to 100. The x-axis lists different metrics (e.g., C11, T1, C5, M2, etc.). Two experiments are compared: Isolation (red crosses), where models are trained solely on one code metric, and Leave-one-out (blue crosses), where models are trained on all but one code metric. The red crosses show performance dropping below 75\% for most metrics, with some going as low as 20\%, while the blue crosses remain near 100\% across all metrics.}
    \label{fig:retrain-best}
\end{figure}
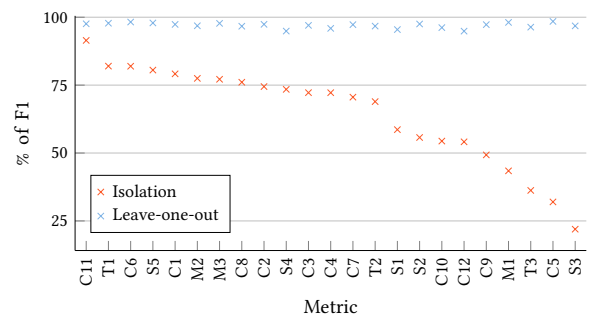

\paragraph{Individual performance.}
As the first experiment, we investigate the performance of individual code metrics on the PrimeVul dataset. To this end, we re-train our classifier on each metric separately and measure their predictive power in isolation. The results of this experiment are presented in \cref{fig:retrain-best}.

We find that the number of local variables (C11) performs best. This single metric alone achieves over 90\% of the F1 score obtained by a model using all code metrics. Notably, it is not the only metric performing well on its own; a total of eight code metrics reach more than 75\% of the combined metrics' performance. We conclude that even basic statistical properties, such as the number of local variables (C11), the number of tree nodes (T1), and the depth of nested control structures (C6), enable achieving performance levels close to state-of-the-art methods for vulnerability discovery.

\paragraph{Leave-one-out performance}
Second, we conduct a leave-one-out experiment, where each metric is excluded once during classifier training. This allows us to assess the significance of each metric and its interplay within the full set. The results of this experiment are also shown in \cref{fig:retrain-best}.

The experiment reveals that missing one individual metric does not cause a significant drop in model performance, indicating that the remaining metrics can compensate for the lost information. As a consequence, none of the code metrics, even those that show high discriminative value in isolation, are crucial for a strong classifier. This result suggests that the observed performance of code metrics is not simply due to a few lucky metrics. Instead, it demonstrates the reliability of the overall discovery approach.

\subsection{Alternative Approaches}

For comparison, we consider static application security testing (SAST) tools and graph neural networks (GNNs) as an additional reference point in our evaluation.

\paragraph{Graph neural networks} We additionally evaluate Devign~\citep{ZhoLiuSioDu+19} and ReVeal~\citep{ChaKriDinRay+22}, two well-established GNNs for function-level vulnerability discovery. Both models are trained for 100 epochs, with early stopping triggered after 10 epochs of no improvement in validation loss. The underlying graph representations are extracted using the tool \emph{Joern}~\citep{joern}. If the tool fails to parse the code, we return an inconclusive prediction with a confidence score of $0.5$. Consistent with the language model experiments, we repeat this experiment 10 times and investigate mean values.

Overall, we find that the GNNs demonstrate significantly lower F1 scores on the PrimeVul dataset with a maximum of 17.8\% ($\pm 3.85$). While their performance is still much better than random guessing, it is only about as good as the language model with the worst performance in our study. Access to the rich representation of code graphs does not prove beneficial in our experiments. Instead, the simple numerical quantities provided through code metrics offer a more effective representation to identify vulnerable code.

\paragraph{Static application security testing} For the SAST tools, we utilize \emph{Rats}~\cite{rats}, one of the earliest freely available examples in this category, and \emph{SemGrep}~\cite{semgrep}, a recent representative. For both, we calculate a weighted sum of warnings and errors for each code sample and apply a threshold to classify them.
With this approach, the maximum F1 score achieved is 14.2\%, placing SAST tools slightly behind GNN-based approaches but still well above general-purpose language models in terms of detection performance.

\section{Root-Cause Analysis}
\label{sec:interpretability}
Why do LLMs for vulnerability discovery fail to outperform simple code metrics? Theoretically, these models have access to a far broader range of information than code metrics could ever provide. LLMs can analyze variable names, inspect structured data types, and track control flow within code, providing a wealth of insights for separating secure from vulnerable code.
However, we observe no significant performance difference between a language model and the classifier trained on code metrics.

We hypothesize that LLMs do not unlock a deeper level of code analysis, despite the information being available. Instead, they rely on basic statistical properties that are equivalent, though not necessarily identical, to those calculated by syntactic code metrics. To support our hypothesis, we identify four essential preconditions and conduct a series of experiments to confirm them:

\begin{enumerate}
    \setlength{\itemsep}{4pt}
    \item[P1] \emph{Information access.} LLMs and code metrics can only rely on the same information if both have access to it. In our first experiment, we explore whether equivalent information is present by reconstructing code metrics from the learning models' embeddings.

    \item[P2] \emph{No cross-information gain.} If LLMs and code metrics use complementary information in their decisions, combining them should enhance performance and invalidate our hypothesis. Therefore, we test whether their combination leads to a measurable performance gain.

    \item[P3] \emph{Prediction correlation.} Even if LLMs and code metrics have access to the same information, they may not use it in the same way during inference. In our third experiment, we therefore measure the correlation between code metrics and the predictions of LLMs.
    
    \item[P4] \emph{Causal dependence.} Correlation only indicates a relationship between the information and the prediction, but not the type of relationship. Thus, to measure the direction of the relationship, we determine the causal dependency between code metrics and language models.

\end{enumerate}

This testing procedure is inspired by the causal hierarchy framework of \citet{Pea09}, encompassing both correlation validation and causal effect inference. While this approach cannot reveal the complex inner workings of LLMs, it helps rule out alternative explanations for their performance similarity to syntactic code metrics. If all preconditions hold, we must conclude that the information LLMs process for vulnerability discovery is not fundamentally different from that provided by code metrics.

\subsection{Information Access (P1)}

In our first experiment, we aim to determine whether LLMs have access to the same information as the code metrics. Recall that the code language models in our evaluation comprise two components: an embedding model and a classification head. Any information used for prediction must be encoded in the embedding before being passed to the classification head. Therefore, if the models have access to data equivalent to code metrics, that information must be present in the embedding. To test this hypothesis, we train linear regression models to predict the code metrics from the language models’ embeddings.

This approach aligns with recent research on the interpretability of large language models. For instance, \citet{PCV+24} advocate for the \textit{linear representation hypothesis}, which states that semantic information is represented linearly as directions in the models' embedding space. This hypothesis supports the use of linear probes to extract concept-level information from the hidden representations of the models~\cite{AlaBen17}.

\paragraph{Experimental setup.}
We consider all language models for vulnerability discovery from the previous section trained on PrimeVul and fit linear regressors on the penultimate layer of these models using the training split. For evaluation, we use the test split, ensuring a temporal difference between the training and evaluation data.

A hindrance to this training is that most code metrics represent counts of pattern occurrences, so their values typically follow a Poisson distribution rather than a Gaussian distribution. To address this, we train the regression models to predict the logarithm of the metric values, as the log-transformed values better approximate a normal distribution~\cite{loglinearRegression}. Furthermore, the metrics are computed only for the respective context of the LLM instead of the complete function. This is because the models lack information beyond their context, making predictions outside of them impossible.

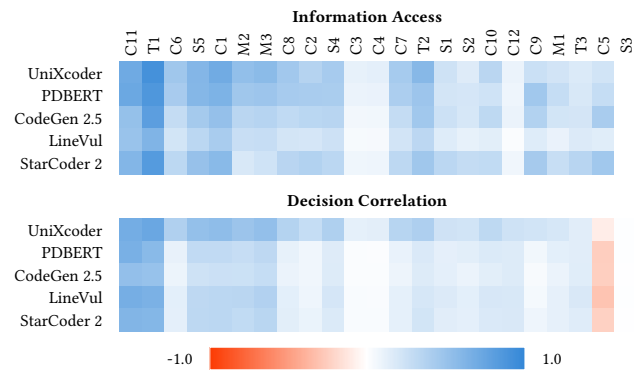
\begin{figure}[b]
    \centering
    \vspace{0.25cm}
%
%
%

\resizebox{\linewidth}{!}{
\begin{tikzpicture}[scale=0.42]
    \foreach \y [count=\n] in {
        {0.6818, 0.7323, 0.3881, 0.5254, 0.5535, 0.4893, 0.5292, 0.3685, 0.2833, 0.3929, 0.1289, 0.1376, 0.3533, 0.3973, 0.2384, 0.2243, 0.3177, 0.2428, 0.2255, 0.2040, 0.1463, -0.0680, 0.0104},
        {0.6607, 0.5789, 0.1153, 0.3049, 0.3045, 0.2801, 0.3219, 0.1190, 0.0725, 0.1628, 0.0244, 0.0219, 0.1066, 0.1815, 0.1297, 0.1415, 0.1762, 0.1663, 0.0708, 0.1375, 0.1463, -0.1854, 0.0091},
        {0.5294, 0.5107, 0.0933, 0.2412, 0.2567, 0.2540, 0.2852, 0.0962, 0.0738, 0.1679, 0.0260, 0.0256, 0.0891, 0.1690, 0.1323, 0.0980, 0.1446, 0.1668, 0.0399, 0.0981, 0.1518, -0.1821, 0.0063},
        {0.6723, 0.6517, 0.1356, 0.3213, 0.3341, 0.3412, 0.3796, 0.1369, 0.0833, 0.2047, 0.0337, 0.0298, 0.1266, 0.2030, 0.1711, 0.1373, 0.1985, 0.1938, 0.0529, 0.1213, 0.1887, -0.2244, 0.0106},
        {0.6168, 0.6033, 0.1328, 0.3243, 0.3326, 0.3032, 0.3448, 0.1405, 0.1013, 0.1771, 0.0344, 0.0321, 0.1267, 0.2203, 0.1489, 0.1309, 0.1835, 0.1889, 0.0823, 0.1215, 0.1569, -0.1741, 0.0059},
    } {
      \foreach \x [count=\m] in \y {
        \pgfmathsetmacro{\twox}{abs(\x)*160}
        \pgfmathsetmacro{\signx}{round((\x/abs(\x) + 1) / 2)}
        \pgfmathsetmacro{\positivex}{\signx}
        \pgfmathsetmacro{\negativex}{1 - \signx}
        \node[fill=poscolor!\twox!white, minimum size=4.4mm, text=white, opacity=\positivex] at (\m,-\n-7) { };
        \node[fill=negcolor!\twox!white, minimum size=4.4mm, text=white, opacity=\negativex] at (\m,-\n-7) { };
        }
    }

    \foreach \a [count=\i] in {C11,T1,C6,S5,C1,M2,M3,C8,C2,S4,C3,C4,C7,T2,S1,S2,C10,C12,C9,M1,T3,C5,S3} {
        \node[minimum size=6mm, rotate=90, anchor=west] at (\i, -0.5) {\a};
    }
    \foreach \a [count=\i] in {UniXcoder,PDBERT,CodeGen 2.5,LineVul,StarCoder 2} {
        \node[minimum size=6mm, xshift=0.0cm, anchor=east] at (0,-\i-7) {\a};
    }

    \foreach \y [count=\n] in {
        {0.7092, 0.9178, 0.4662, 0.6072, 0.6967, 0.5358, 0.5702, 0.4581, 0.3644, 0.4306, 0.1160, 0.1353, 0.4367, 0.5882, 0.2401, 0.1640, 0.3390, 0.0975, 0.2593, 0.2218, 0.1787, 0.2240, 0.0001},
        {0.7286, 0.8652, 0.4304, 0.5996, 0.6374, 0.4655, 0.4833, 0.4324, 0.4231, 0.4177, 0.0982, 0.1042, 0.4102, 0.4836, 0.2142, 0.2063, 0.2349, 0.0858, 0.4619, 0.2870, 0.1899, 0.2848, 0.0021},
        {0.5208, 0.8045, 0.2902, 0.4441, 0.5223, 0.3416, 0.3574, 0.3087, 0.3371, 0.3473, 0.0644, 0.0705, 0.2909, 0.4610, 0.2528, 0.2019, 0.3176, 0.1006, 0.3776, 0.2207, 0.2131, 0.4218, 0.0001},
        {0.5041, 0.6254, 0.2193, 0.3324, 0.4223, 0.2719, 0.2829, 0.2165, 0.2077, 0.2474, 0.0447, 0.0466, 0.2221, 0.3307, 0.1612, 0.1171, 0.1435, 0.0255, 0.1614, 0.1045, 0.1719, 0.1510, 0.0004},
        {0.6107, 0.8357, 0.3289, 0.5131, 0.5787, 0.1901, 0.2364, 0.3464, 0.3773, 0.3353, 0.0775, 0.0827, 0.3269, 0.4691, 0.3333, 0.2885, 0.3041, 0.0742, 0.4393, 0.2779, 0.3477, 0.4613, 0.0001},
    } {
      \foreach \x [count=\m] in \y {
        \pgfmathsetmacro{\positivetwox}{abs(\x)*160}
        \pgfmathsetmacro{\negativetwox}{abs(\x)*120}
        \pgfmathsetmacro{\signx}{round((\x/abs(\x) + 1) / 2)}
        \pgfmathsetmacro{\positivex}{\signx}
        \pgfmathsetmacro{\negativex}{1 - \signx}
        \node[fill=poscolor!\positivetwox!white, minimum size=4.4mm, text=white, opacity=\positivex] at (\m,-\n) { };
        \node[fill=negcolor!\negativetwox!white, minimum size=4.4mm, text=white, opacity=\negativex] at (\m,-\n) { };
        }
    }

    \foreach \a [count=\i] in {UniXcoder,PDBERT,CodeGen 2.5,LineVul,StarCoder 2} {
        \node[minimum size=6mm, xshift=0.0cm, anchor=east] at (0,-\i) {\a};
    }

    \node[anchor=south] at (11.5, 1) {\textbf{Information Access}};
    \node[anchor=south] at (11.5, -7.2) {\textbf{Decision Correlation}};
  
    \draw[left color=negcolor!120!white, minimum size=6mm, right color=white, color=white] (4.5, -13.2) rectangle (11.5, -14.2);
    \draw[left color=white, minimum size=6mm, right color=poscolor!160!white, color=white] (11.5, -13.2) rectangle (18.5, -14.2);
    
    \node[anchor=east] at (4, -13.7) {-1.0};
    \node[anchor=west] at (19, -13.7) {1.0};
\end{tikzpicture}}
    \caption{The coefficient of determination for the prediction and the correlation coefficient to the predicted label for each feature. The features are sorted by the mutual information between the feature and the ground truth label with more predictive features on the left.}
    \Description{Two heatmaps compare features across different models. The top heatmap, labeled Information Access, shows the coefficient of determination values for features across models UniXcoder, PDBERT, CodeGen 2.5, LineVul, and StarCoder 2. The x-axis lists features such as C11, T1, C6, S1, M2, and others. Darker blue cells indicate higher coefficients. The bottom heatmap, labeled Decision Correlation, shows the correlation coefficient of features to the predicted label for the same models and features. The color scale ranges from -1.0 in red to 1.0 in blue, with most values in light blue, some few in darker shades of blue and one showing red shades on the right. Both heatmaps show a roughly similar pattern for all models.}
    \label{fig:metric-correlation2}
\end{figure}

\paragraph{Results.} The results of this experiment are summarized in \cref{fig:metric-correlation2}, which presents the coefficient of determination for each learned metric. A higher coefficient indicates a better ability to predict the corresponding code metric. The data indicates that most metrics can be effectively learned. Simpler metrics, such as the number of local variables~(C11), the number of if statements without else clauses~(S5), and the tree height~(T2), are among the best learned. Notably, the metric with the highest performance is the number of nodes~(T1), with a coefficient of determination of 0.91 for UniXcoder, 0.86 for PDBERT, and 0.8 for CodeGen 2.5. Additionally, some complex features, including the control structure complexity~(C8) and the count of pointer arithmetic operations~(M3), are also learned with high accuracy, demonstrating the models' capability to capture intricate patterns.

\subsection{No Cross-Information Gain (P2)}

In our second experiment, we aim to demonstrate that the information processed by LLMs is not complementary to code metrics. If it were, combining both should improve the performance of vulnerability discovery and invalidate our hypothesis about their relationship. Conversely, if incorporating code metrics into an LLM does not lead to performance gains, this suggests that the relevant information is already present in identical or equivalent form.

\paragraph{Experimental setup.}
We concatenate the feature vector of the 23 syntactic metrics with the model embeddings from the penultimate layer and retrain the classification heads for all language models using the PrimeVul dataset. We then measure the performance difference between models augmented with code metrics and those without across 10 runs.

\paragraph{Results.}
We find that the average difference in F1-scores is less than 0.14 percentage points, which is substantially smaller than the variance observed between two independently trained instances of the same model. This result indicates that the code metrics do not provide additional information beyond what is already captured by the language model, suggesting that the available information overlaps rather than being complementary.

\subsection{Prediction Correlation (P3)}

We proceed to investigate whether LLMs and code metrics not only have access to the same information but also rely on it for identifying vulnerabilities. To this end, we analyze the correlation between individual code metrics and the predictions of LLMs. In the framework of \citet{Pea09}, this correlation is an essential prerequisite for demonstrating causal dependence.

\paragraph{Experimental setup.}
Specifically, we compute the Pearson correlation coefficient between each syntactic code metric and the predictions of the LLMs on the PrimeVul dataset, averaging the coefficients over ten experimental runs.

\paragraph{Results.}
The results are presented in the lower half of \autoref{fig:metric-correlation2}. We observe that all models exhibit correlation with the code metrics, where the strength depends on the accessibility of the information. A weaker correlation is observed when the information is inaccessible, whereas accessible information leads to a strong correlation. This suggests that the information available to the models is also used for predictions. Interestingly, the metric C5 (number of parameters) shows a negative correlation, indicating that the models treat it as a negative predictor of vulnerabilities.

\subsection{Causal Dependence (P4)}
So far, we have confirmed that LLMs have access to information equivalent to code metrics and that their predictions are strongly correlated with them. However, these conclusions are solely based on passive observations. Establishing a causal relationship requires demonstrating the direct impact of code metrics on predictions through an \emph{interventional approach}. This is analogous to a dose-response analysis in epidemiological studies~\citep{Hil+65}. For instance, if cigarette consumption not only correlates with the development of lung cancer, but also an increase in smoking leads to a higher risk of the disease, a causal effect is likely.

\paragraph{Experimental setup.}
For our investigation, we formulate causal dependency as follows: If an LLM depends on code metrics for its decisions, then changes in the metrics should systematically influence its predictions. To test this condition, we require a method for modifying code along with its associated metrics and a measure to quantify the influence of this change on LLMs. 

\paragraph{(a) Intervening metrics.}   
Ideally, causal dependency is measured by modifying only a single variable, such as a specific metric. However, since code metrics are inherently dependent on the underlying code, it is impossible to change one metric in isolation without affecting others. Additionally, artificially altering code risks creating unnatural samples that fall outside the distribution learned by the models.
To ensure that code modifications remain both natural and minimal, we retrieve the previous and subsequent commits for each code sample in PrimeVul from their respective Git repositories. We then apply the corresponding patches to each function, thereby intervening on the code metrics in a realistic manner.

\paragraph{(b) Measuring dependency.}

Given a piece of code $c$ and its modifications, we can compute two quantities: the difference in predictions of an LLM, denoted as $\Delta y \in \mathbb{R}$, and the difference in code metrics, $\Delta \mu \in \mathbb{R}^{23}$. Since these quantities operate on entirely different scales, we aggregate the metric differences $\Delta \mu$ into a single value $\Delta s \in \mathbb{R}$ by training a linear regression model to predict $\Delta y$ from $\Delta \mu$. 
We then compare the aggregation $\Delta s$ with $\Delta y$ using the \emph{coefficient of determination}~$R^2$, a standard measure in statistics for dependence~\citep{Wri+21}. In our case, the coefficient can be defined as
\[
R^2 = 1 - \frac{\sum_{\code} \left ( 
\Delta y_c - \Delta s_c
\right )^2 }%
{\sum_{\code} \left ( 
\Delta y_c - \Delta \bar{y}
\right )^2  }
\]
where $\Delta \bar{y}$ is the average of prediction differences.
Intuitively, $R^2$ represents the proportion of variance in the prediction differences that can be explained solely by code metrics. If changes in the code lead to identical variations in both the LLM predictions and the metrics, then $R^2 = 1$. Conversely, if the changes are indistinguishable from the average of all differences, then $R^2 = 0$.

\paragraph{Results.} The code metrics achieve an $R^2$ of 0.42 for UniXcoder, 0.38 for PDBERT, and 0.25 for LineVul, indicating a notable causal dependency. In line with our hypothesis, the strongest dependency is observed for the LLM that achieves the best performance. While the LLMs do not exclusively rely on code metrics for finding vulnerabilities, the metrics account for between 20\% and 40\% of information in their predictions. Note that these results likely underestimate the true dependency, as the aggregation $\Delta s$ is based on a linear regression, whereas the actual relationship between the metrics and the predictions is likely more complex.

For the larger models, CodeGen 2.5 and StarCoder 2, we measure $R^2$ close to 0, indicating no significant dependence. This suggests that either the larger models perform more sophisticated analyses or that our set of 23 metrics misses simple features used by the models. The evaluation results of the larger models support the latter explanation, as more sophisticated analysis should lead to significantly better performance in the vulnerability discovery task. Furthermore, recent studies have shown that models with higher dimensionality exhibit more differentiated and specific features, which may be responsible for their predictions~\cite{bricken2023monosemanticity}.

\subsection{Summary}

\begin{table}[b]
    \caption{Summary of the results of the root-cause analysis.}
    \label{tab:root-cause-results}
    \centering
    \resizebox{.85\linewidth}{!}{
    \small
\begin{tabular}{l r >{\centering\arraybackslash}p{2em} >{\centering\arraybackslash}p{2em} >{\centering\arraybackslash}p{2em} >{\centering\arraybackslash}p{2em} >{\centering\arraybackslash}p{2em} >{\centering\arraybackslash}p{2em}}
    \toprule 
    \textbf{Model} & \multicolumn{1}{c}{\textbf{Parameters}} \; & \textbf{P1} & \textbf{P2} &  \textbf{P3} & \textbf{P4} \\
    \midrule
    UniXcoder   & $1.25 \times 10^8$ \; & \cmark & \cmark & \cmark & \cmark \\
    PDBERT      & $1.25 \times 10^8$ \; & \cmark & \cmark & \cmark & \cmark \\
    LineVul     & $1.25 \times 10^8$ \; & \cmark & \cmark & \cmark & \cmark \\
    CodeGen 2.5 & $6.69 \times 10^9$ \; & \cmark & \cmark & \cmark & \xmark \\
    StarCoder 2 & $7.17 \times 10^9$ \; & \cmark & \cmark & \cmark & \xmark \\
    \bottomrule
\end{tabular}

    }
\end{table}

Our root-cause analysis, summarized in \cref{tab:root-cause-results}, reveals that all evaluated models have access to information provided by code metrics, exhibit correlation with them, and do not benefit from cross-information. These findings indicate that LLMs predominantly rely on information equivalent but not necessarily identical to the considered syntactic code metrics. For medium-sized LLMs, we identify a clear causal dependency, quantifying the extent to which their predictions are influenced by code metrics. However, this causal indicator is absent in larger models, suggesting that they correlate with code metrics but may rely on alternative information for making their decisions.

\section{Discussion \& Recommendations}
\label{sec:discussion}
Our investigation leads to a disappointing outcome: despite the impressive capabilities of language models in other domains, their performance in vulnerability discovery is not significantly different from that of a simple baseline. The substantial resources required to train these models, along with the considerable effort in curating high-quality training datasets, do not yield a substantial advantage over simple techniques developed decades ago.

These findings are not entirely unexpected. Previous studies have highlighted limitations of machine learning for the task of vulnerability discovery, pointing to issues such as inappropriate benchmarks and low predictive performance~\citep{ArpQuiPenWar+22, ChaKriDinRay+22, DinFuIbrSit+25, JimRwePapSar+19, RisBöh24}. Furthermore, the recent study by \citet{DinFuIbrSit+25} indicates that scaling language models further up is unlikely to address this issue, as larger models do not automatically perform better \cb{on this task}. Unfortunately, this generally creates a rather pessimistic outlook. Current research appears to have reached a plateau, and simply expanding the amount of training data, scaling up learning models, or making incremental adjustments does not seem to offer a particularly promising path forward at this stage.

Therefore, we suggest taking a step back to leverage the insights from our analysis in developing new directions \cb{unlocking more promising vulnerability discovery approaches}. In the following, we summarize these insights into actionable recommendations:

\paragraph{R1: Code metrics are relevant baselines.} Despite their simplicity, classifiers using code metrics should be employed as baselines for vulnerability discovery. If we expect modern learning models to uncover complex patterns in code, it is essential to contrast them with simple approaches. Hence, we recommend using classifiers based on code metrics as a sanity check in empirical evaluations. It is worth noting that our approach is automatically optimized and trained within 10 minutes on PrimeVul, adding minimal overhead as a baseline in experimentation.

\paragraph{R2: Occam's razor matters.} At first glance, any improvement in vulnerability discovery appears beneficial. However, when comparing different approaches, it is crucial to balance model complexity against detection capabilities~\citep{HasTibFri09}. For instance, in our experiments in \cref{sec:evaluation}, UniXcoder--if at all--only marginally outperforms the classifier based on code metrics. Yet, our model uses 94\% fewer parameters than the language model and runs without specialized hardware, greatly enhancing efficiency and facilitating integration into development workflows. Therefore, we recommend to evaluate model performance relative to the model size. Ideally, a model's capabilities should be plotted as a function of its number of parameters or a similarly meaningful measure that considers both model complexity and performance.

\paragraph{R3: Rethinking learning-based discovery.} Current research largely approaches the task of identifying vulnerable code as a black-box problem on input samples with limited context information, where learning models are expected to achieve high performance from scratch. The success of general pre-training has further reinforced this perspective, making it challenging to discern whether a model is truly learning patterns of vulnerabilities or merely replicating simple metrics.

To address this issue, we suggest rethinking the application of machine learning in vulnerability discovery:
\begin{enumerate}

\setlength{\itemsep}{4pt}

\item \emph{Better code representations: } It is unclear whether current representations, such as token sequences and code graphs~(for GNNs), provide a suitable basis for learning patterns of vulnerabilities. Based on our findings, there is no evidence that these representations on their own unlock a deeper level for code inspection within the learning models.

\item \emph{Improvements that matter: } The sole reliance on performance-driven loss functions may limit a model's ability to improve beyond current methods. Ultimately, we are interested in spotting those kinds of defects that remain undetected so far. One strategy could involve making the employed loss function aware of this objective \citep{ClaYatZet19} by incorporating feedback from other approaches for vulnerability discovery.

\item \emph{Steps rather than leaps: } There is increasing evidence that vulnerability discovery cannot be approached as an end-to-end learning task~\citep{DinFuIbrSit+25}. New intermediate representations and learning steps are likely essential for improving performance and enforcing the required deeper insights into the code that existing approaches as well as code metrics cannot offer.

\item \emph{Better vulnerability benchmarks:} Lastly, we need to move beyond the artificial setting of detecting vulnerabilities from individual functions alone. As \citet{RisBoe24} point out, many vulnerabilities used in benchmarks cannot be reliably identified without additional context. That is, code metrics also perform well because essential context, beneficial for more complex models, is simply not available.

\end{enumerate}

To foster this development and further advancements in the field of vulnerability discovery, we make our implementation and experimental framework publicly available\footnote{\url{https://github.com/mlsec-group/cheetah}}.

\section{Related Work}
\label{sec:related-work}
Critical reflections have a long tradition in security-related research, especially when machine learning techniques are used to address challenging problems. Notable examples of this line of research include critical reviews of machine learning in network intrusion detection~\citep{SomPax10, FloEngAspDes+24}, website fingerprinting~\citep{JuaAfrAcaDía+14, CheJanTro22}, and security applications in general~\citep{ArpQuiPenWar+22}. We continue this line of work by offering a critical reflection on vulnerability discovery through the lens of code metrics, building on prior research that examines data quality and detection capabilities.

\paragraph{Reflections on data quality.}
Several previous studies have focused on understanding and improving data quality in learning-based approaches to vulnerability discovery. As one of the first, \citet{JimRwePapSar+19} propose an improved labeling scheme that accounts for temporal dependencies between vulnerabilities, helping to prevent potential data leaks from other splitting approaches. 
Building on this, \citet{ChaKriDinRay+22} analyze the quality of training datasets, focusing on the sources of labeling and the naturalness of source code samples. They contribute an additional, more realistic dataset and propose a representation learning and sampling technique to fight the strong class imbalance. 

Follow-up work by \citet{SejDasShaMed24} and \citet{RisBoe24} investigate the quality of code samples, revealing that many vulnerabilities cannot be detected from these samples alone. This research provides evidence that detecting a vulnerability often requires assessing multiple parts of a software in combination and consequently, is often infeasible without further context information.
Constructing suitable and realistic datasets not only faces the issue of sample quality but a fundamental trade-off between incorporating relevant context and restricting the input size to a manageable amount.
Most recently, \citet{DinFuIbrSit+25} further enhance the label accuracy of existing benchmark datasets on function-level granularity, resulting in the curation of the PrimeVul dataset that we used in our study.

\paragraph{Reflections on detection capabilities.}
Another direction of prior work has focused on investigating the capabilities and limitations of learning-based approaches to vulnerability discovery.
For example, it has been shown that language models tend to overfit to code variants that are present in the training data, prompting the development of augmentation techniques to mitigate this overfitting by applying code transformations during training that do not semantically alter the code samples~\citep{RisBöh24,brokenPromises}. Additionally, \citet{UllHanPujPea+24} demonstrate that language models produce non-deterministic responses and often provide incorrect or unfaithful reasoning over vulnerable code. Similarly, studies on limitations of graph neural networks describe effects like over-squashing~\citep{AloYah2021}, over-smoothing~\citep{Ham2020} and low performances on heterophilic graphs~\citep{PlaKuzDisBab+23}, such as those used in vulnerability discovery.

\medskip
In contrast to this previous work, our study aims to better understand the low performance of current learning models. By analyzing them through the lens of code metrics, we reveal their tendency to focus on basic statistical properties rather than code analysis. This finding aligns with observations of limited robustness and unclear reasoning from prior studies, adding a new facet to our understanding of why learning models continue to struggle with identifying vulnerabilities.

\section{Limitations}
\label{sec:limitations}
Our retrospective analysis of LLMs for vulnerability discovery inherently carries limitations due to its empirical nature. Potential threats to validity arise from the evaluation data as well as the choice of syntactic code metrics used in our experiments and the general possibility of confounding variables in machine learning, which we discuss in the following.

\paragraph{Dataset quality.} The quantity and quality of data for machine learning is a critical factor in vulnerability discovery and remains an active research area~\citep{CheDinAloChe+23, DinFuIbrSit+25, ArpQuiPenWar+22, ChaKriDinRay+22}. Our findings are tied to the characteristics of the employed datasets. While LLMs could potentially outperform code metrics if considerably larger and more representative datasets were available for evaluation, with PrimeVul we are working with the currently largest and most refined dataset in this domain. PrimeVul is the result of substantial efforts within the research community, and there is no indication that significantly larger datasets will become available in the near future due to the difficulty of labeling security defects automatically. 

\clearpage
\paragraph{Selection of metrics.} For our experiments, we consider a set of 23 syntactic code metrics. These metrics include re-implementations from prior work on identifying vulnerable code regions, as well as newly designed metrics aimed at capturing common characteristics of security flaws. We made a best effort to compile a comprehensive and representative set of metrics derived from syntax. However, this selection naturally does not encompass the full range of metrics available in software engineering, which also includes dynamic and operational measures. Some of these additional metrics may introduce new information and could correlate even more strongly with recent LLMs than the ones we considered. Nevertheless, this does not invalidate our findings; rather, it demonstrates that our results establish a lower bound on the predictive power that can be explained through the lens of code metrics.

\cbstart
\paragraph{Hidden variables.} As in most machine learning research, hidden variables that influence predictions cannot be fully ruled out. For instance, a code change may alter semantics without affecting code metrics, in which case the metrics cannot fully account for the change in model prediction. This could stem from hidden variables exploited by the models or from further code metrics not included in our assessment. Likewise, a confounding hidden variable might affect both the metrics and the prediction outcome. In the first case, our evaluation likely underestimates the impact of code metrics, while in the second it overestimates their influence.
\cbend

\section{Conclusion}
\label{sec:conclusion}
Our study uncovers an unexpected behavior of current LLMs for vulnerability discovery. While these models are widely believed to offer sophisticated analysis, we find that they largely capture basic statistical properties rather than deeper, structural insights of code. Simple code metrics can measure the very same properties using just a fraction of the computing resources. In combination with a traditional classifier, they currently perform on par with the best vulnerability discovery models. Our root-cause analysis confirms that the models closely align with code metrics, drawing on equivalent information for their decisions. The best-performing LLMs even exhibit a causal dependency on them.

While code metrics are valuable for identifying potential issues, they are, by design, insufficient for accurately pinpointing vulnerabilities. This limitation is evident from the consistently low performance of all methods in our experiments. Consequently, we do not advocate code metrics as a promising direction for advancing vulnerability discovery. Instead, we propose to take a step back and use code metrics as a reality check for rethinking the process of vulnerability discovery. We hope that the recommendations outlined in this work contribute to the development of more effective methods capable of obtaining deeper insights into code when learning and identifying security flaws.

\section*{Acknowledgments}
\label{sec:acknowledgments}
\cbstart
This work was supported by the European Research Council (ERC) under the consolidator grant MALFOY (101043410), the German Federal Ministry of Research, Technology and Space under the grant AIgenCY (16KIS2012), and the Deutsche Forschungsgemeinschaft~(DFG, German Research Foundation) under Germany's Excellence Strategy -- EXC 2092 CASA (390781972).
\cbend

\balance
\bibliography{
    bib/acsac.bib,
    bib/asiaccs.bib,
    bib/ccs.bib,
    bib/raid.bib,
    bib/eurosp.bib,
    bib/icse.bib,
    bib/ndss.bib,
    bib/sp.bib,
    bib/uss.bib,
    bib/pldi.bib,
    bib/nips.bib,
    bib/additional.bib
}


\begin{thebibliography}{70}


\ifx \showCODEN    \undefined \def \showCODEN     #1{\unskip}     \fi
\ifx \showISBNx    \undefined \def \showISBNx     #1{\unskip}     \fi
\ifx \showISBNxiii \undefined \def \showISBNxiii  #1{\unskip}     \fi
\ifx \showISSN     \undefined \def \showISSN      #1{\unskip}     \fi
\ifx \showLCCN     \undefined \def \showLCCN      #1{\unskip}     \fi
\ifx \shownote     \undefined \def \shownote      #1{#1}          \fi
\ifx \showarticletitle \undefined \def \showarticletitle #1{#1}   \fi
\ifx \showURL      \undefined \def \showURL       {\relax}        \fi
\providecommand\bibfield[2]{#2}
\providecommand\bibinfo[2]{#2}
\providecommand\natexlab[1]{#1}
\providecommand\showeprint[2][]{arXiv:#2}

\bibitem[Alain and Bengio(2017)]%
        {AlaBen17}
\bibfield{author}{\bibinfo{person}{Guillaume Alain} {and} \bibinfo{person}{Yoshua Bengio}.} \bibinfo{year}{2017}\natexlab{}.
\newblock \showarticletitle{Understanding intermediate layers using linear classifier probes}. In \bibinfo{booktitle}{\emph{Proc. of International Conference on Learning Representations (ICLR)}}.
\newblock


\bibitem[Alon and Yahav(2021)]%
        {AloYah2021}
\bibfield{author}{\bibinfo{person}{Uri Alon} {and} \bibinfo{person}{Eran Yahav}.} \bibinfo{year}{2021}\natexlab{}.
\newblock \showarticletitle{On the {{Bottleneck}} of {{Graph Neural Networks}} and Its {{Practical Implications}}}. In \bibinfo{booktitle}{\emph{Proc. of the International Conference on Learning Representations (ICLR)}}.
\newblock


\bibitem[Arp et~al\mbox{.}(2022)]%
        {ArpQuiPenWar+22}
\bibfield{author}{\bibinfo{person}{Daniel Arp}, \bibinfo{person}{Erwin Quiring}, \bibinfo{person}{Feargus Pendlebury}, \bibinfo{person}{Alexander Warnecke}, \bibinfo{person}{Fabio Pierazzi}, \bibinfo{person}{Christian Wressnegger}, \bibinfo{person}{Lorenzo Cavallaro}, {and} \bibinfo{person}{Konrad Rieck}.} \bibinfo{year}{2022}\natexlab{}.
\newblock \showarticletitle{Dos and Don'ts of Machine Learning in Computer Security}. In \bibinfo{booktitle}{\emph{Proc. of the USENIX Security Symposium}}.
\newblock


\bibitem[Backes et~al\mbox{.}(2017)]%
        {BacRieSkoSto+17}
\bibfield{author}{\bibinfo{person}{Michael Backes}, \bibinfo{person}{Konrad Rieck}, \bibinfo{person}{Malte Skoruppa}, \bibinfo{person}{Ben Stock}, {and} \bibinfo{person}{Fabian Yamaguchi}.} \bibinfo{year}{2017}\natexlab{}.
\newblock \showarticletitle{Efficient and Flexible Discovery of PHP Application Vulnerabilities}. In \bibinfo{booktitle}{\emph{Proc. of the IEEE European Symposium on Security and Privacy (EuroS\&P)}}.
\newblock


\bibitem[Bhandari et~al\mbox{.}(2021)]%
        {BhaNasMoo21}
\bibfield{author}{\bibinfo{person}{Guru~Prasad Bhandari}, \bibinfo{person}{Amara Naseer}, {and} \bibinfo{person}{Leon Moonen}.} \bibinfo{year}{2021}\natexlab{}.
\newblock \showarticletitle{{CVEfixes:} {A}utomated collection of vulnerabilities and their fixes from open-source software}. In \bibinfo{booktitle}{\emph{Proc. of the International Conference on Predictive Models and Data Analytics in Software Engineering (PROMISE)}}.
\newblock


\bibitem[Bricken et~al\mbox{.}(2023)]%
        {bricken2023monosemanticity}
\bibfield{author}{\bibinfo{person}{Trenton Bricken}, \bibinfo{person}{Adly Templeton}, \bibinfo{person}{Joshua Batson}, \bibinfo{person}{Brian Chen}, \bibinfo{person}{Adam Jermyn}, \bibinfo{person}{Tom Conerly}, \bibinfo{person}{Nick Turner}, \bibinfo{person}{Cem Anil}, \bibinfo{person}{Carson Denison}, \bibinfo{person}{Amanda Askell}, \bibinfo{person}{Robert Lasenby}, \bibinfo{person}{Yifan Wu}, \bibinfo{person}{Shauna Kravec}, \bibinfo{person}{Nicholas Schiefer}, \bibinfo{person}{Tim Maxwell}, \bibinfo{person}{Nicholas Joseph}, \bibinfo{person}{Zac Hatfield-Dodds}, \bibinfo{person}{Alex Tamkin}, \bibinfo{person}{Karina Nguyen}, \bibinfo{person}{Brayden McLean}, \bibinfo{person}{Josiah~E Burke}, \bibinfo{person}{Tristan Hume}, \bibinfo{person}{Shan Carter}, \bibinfo{person}{Tom Henighan}, {and} \bibinfo{person}{Christopher Olah}.} \bibinfo{year}{2023}\natexlab{}.
\newblock \showarticletitle{Towards Monosemanticity: Decomposing Language Models With Dictionary Learning}.
\newblock \bibinfo{journal}{\emph{Transformer Circuits Thread}} (\bibinfo{year}{2023}).
\newblock


\bibitem[Chakraborty et~al\mbox{.}(2022)]%
        {ChaKriDinRay+22}
\bibfield{author}{\bibinfo{person}{Saikat Chakraborty}, \bibinfo{person}{Rahul Krishna}, \bibinfo{person}{Yangruibo Ding}, {and} \bibinfo{person}{Baishakhi Ray}.} \bibinfo{year}{2022}\natexlab{}.
\newblock \showarticletitle{Deep Learning Based Vulnerability Detection: Are We There Yet?}
\newblock \bibinfo{journal}{\emph{{IEEE} Transactions Software Engineering}} \bibinfo{volume}{48}, \bibinfo{number}{9} (\bibinfo{year}{2022}), \bibinfo{pages}{3280--3296}.
\newblock


\bibitem[Chen et~al\mbox{.}(2023)]%
        {CheDinAloChe+23}
\bibfield{author}{\bibinfo{person}{Yizheng Chen}, \bibinfo{person}{Zhoujie Ding}, \bibinfo{person}{Lamya Alowain}, \bibinfo{person}{Xinyun Chen}, {and} \bibinfo{person}{David Wagner}.} \bibinfo{year}{2023}\natexlab{}.
\newblock \showarticletitle{DiverseVul: A New Vulnerable Source Code Dataset for Deep Learning Based Vulnerability Detection}. In \bibinfo{booktitle}{\emph{Proc. of the International Symposium on Research in Attacks, Intrusions and Defenses (RAID)}}. \bibinfo{pages}{654--668}.
\newblock


\bibitem[Cheng et~al\mbox{.}(2021)]%
        {CheWanHuaXu+21}
\bibfield{author}{\bibinfo{person}{Xiao Cheng}, \bibinfo{person}{Haoyu Wang}, \bibinfo{person}{Jiayi Hua}, \bibinfo{person}{Guoai Xu}, {and} \bibinfo{person}{Yulei Sui}.} \bibinfo{year}{2021}\natexlab{}.
\newblock \showarticletitle{{DeepWukong}: {S}tatically Detecting Software Vulnerabilities Using Deep Graph Neural Network}.
\newblock \bibinfo{journal}{\emph{ACM Transactions on Software Engineering and Methodology}} \bibinfo{volume}{30}, \bibinfo{number}{3} (\bibinfo{year}{2021}).
\newblock


\bibitem[Cherubin et~al\mbox{.}(2022)]%
        {CheJanTro22}
\bibfield{author}{\bibinfo{person}{Giovanni Cherubin}, \bibinfo{person}{Rob Jansen}, {and} \bibinfo{person}{Carmela Troncoso}.} \bibinfo{year}{2022}\natexlab{}.
\newblock \showarticletitle{Online Website Fingerprinting: Evaluating Website Fingerprinting Attacks on Tor in the Real World}. In \bibinfo{booktitle}{\emph{Proc. of the USENIX Security Symposium}}. \bibinfo{pages}{753--770}.
\newblock


\bibitem[Clark et~al\mbox{.}(2019)]%
        {ClaYatZet19}
\bibfield{author}{\bibinfo{person}{Christopher Clark}, \bibinfo{person}{Mark Yatskar}, {and} \bibinfo{person}{Luke Zettlemoyer}.} \bibinfo{year}{2019}\natexlab{}.
\newblock \showarticletitle{Don{'}t Take the Easy Way Out: Ensemble Based Methods for Avoiding Known Dataset Biases}. In \bibinfo{booktitle}{\emph{Proc. of the Conference on Empirical Methods in Natural Language Processing~(EMNLP)}}.
\newblock


\bibitem[Coker and Hafiz(2013)]%
        {CokHaf13}
\bibfield{author}{\bibinfo{person}{Zack Coker} {and} \bibinfo{person}{Munawar Hafiz}.} \bibinfo{year}{2013}\natexlab{}.
\newblock \showarticletitle{Program transformations to fix C integers}. In \bibinfo{booktitle}{\emph{Proc. of the International Conference on Software Engineering (ICSE)}}. \bibinfo{pages}{792--801}.
\newblock


\bibitem[Dietz et~al\mbox{.}(2012)]%
        {DieLiRegAdv+12}
\bibfield{author}{\bibinfo{person}{Will Dietz}, \bibinfo{person}{Peng Li}, \bibinfo{person}{John Regehr}, {and} \bibinfo{person}{Vikram~S. Adve}.} \bibinfo{year}{2012}\natexlab{}.
\newblock \showarticletitle{Understanding integer overflow in C/C++}. In \bibinfo{booktitle}{\emph{Proc. of the International Conference on Software Engineering (ICSE)}}.
\newblock


\bibitem[Ding et~al\mbox{.}(2025)]%
        {DinFuIbrSit+25}
\bibfield{author}{\bibinfo{person}{Yangruibo Ding}, \bibinfo{person}{Yanjun Fu}, \bibinfo{person}{Omniyyah Ibrahim}, \bibinfo{person}{Chawin Sitawarin}, \bibinfo{person}{Xinyun Chen}, \bibinfo{person}{Basel Alomair}, \bibinfo{person}{David~A. Wagner}, \bibinfo{person}{Baishakhi Ray}, {and} \bibinfo{person}{Yizheng Chen}.} \bibinfo{year}{2025}\natexlab{}.
\newblock \showarticletitle{Vulnerability Detection with Code Language Models: How Far Are We?\nopunct}. In \bibinfo{booktitle}{\emph{Proc. of the International Conference on Software Engineering (ICSE)}}.
\newblock


\bibitem[Dor et~al\mbox{.}(2003)]%
        {DorRodSag03}
\bibfield{author}{\bibinfo{person}{Nurit Dor}, \bibinfo{person}{Michael Rodeh}, {and} \bibinfo{person}{Shmuel Sagiv}.} \bibinfo{year}{2003}\natexlab{}.
\newblock \showarticletitle{CSSV: towards a realistic tool for statically detecting all buffer overflows in C}. In \bibinfo{booktitle}{\emph{Proc. of the ACM Conference on Programming Language Design and Implementation (PLDI)}}. \bibinfo{pages}{155--167}.
\newblock


\bibitem[Du et~al\mbox{.}(2019)]%
        {DuCheLiGuo+19}
\bibfield{author}{\bibinfo{person}{Xiaoning Du}, \bibinfo{person}{Bihuan Chen}, \bibinfo{person}{Yuekang Li}, \bibinfo{person}{Jianmin Guo}, \bibinfo{person}{Yaqin Zhou}, \bibinfo{person}{Yang Liu}, {and} \bibinfo{person}{Yu Jiang}.} \bibinfo{year}{2019}\natexlab{}.
\newblock \showarticletitle{Leopard: Identifying Vulnerable Code for Vulnerability Assessment through Program Metrics}. In \bibinfo{booktitle}{\emph{Proc. of the International Conference on Software Engineering (ICSE)}}. \bibinfo{pages}{60--71}.
\newblock


\bibitem[Fan et~al\mbox{.}(2020)]%
        {FanLiWanNgu20}
\bibfield{author}{\bibinfo{person}{Jiahao Fan}, \bibinfo{person}{Yi Li}, \bibinfo{person}{Shaohua Wang}, {and} \bibinfo{person}{Tien~N. Nguyen}.} \bibinfo{year}{2020}\natexlab{}.
\newblock \showarticletitle{A {C/C++} Code Vulnerability Dataset with Code Changes and {CVE} Summaries}. In \bibinfo{booktitle}{\emph{Proc. of the International Conference on Mining Software Repositories (MSR)}}.
\newblock


\bibitem[Feng et~al\mbox{.}(2020)]%
        {FenGuoTanDua+20}
\bibfield{author}{\bibinfo{person}{Zhangyin Feng}, \bibinfo{person}{Daya Guo}, \bibinfo{person}{Duyu Tang}, \bibinfo{person}{Nan Duan}, \bibinfo{person}{Xiaocheng Feng}, \bibinfo{person}{Ming Gong}, \bibinfo{person}{Linjun Shou}, \bibinfo{person}{Bing Qin}, \bibinfo{person}{Ting Liu}, \bibinfo{person}{Daxin Jiang}, {and} \bibinfo{person}{Ming Zhou}.} \bibinfo{year}{2020}\natexlab{}.
\newblock \showarticletitle{{CodeBERT:} {A} Pre-Trained Model for Programming and Natural Languages}. In \bibinfo{booktitle}{\emph{Proc. of the Conference on Empirical Methods in Natural Language Processing (EMNLP)}}.
\newblock


\bibitem[Feurer et~al\mbox{.}(2022)]%
        {FeuEggFalLin+22}
\bibfield{author}{\bibinfo{person}{Matthias Feurer}, \bibinfo{person}{Katharina Eggensperger}, \bibinfo{person}{Stefan Falkner}, \bibinfo{person}{Marius Lindauer}, {and} \bibinfo{person}{Frank Hutter}.} \bibinfo{year}{2022}\natexlab{}.
\newblock \showarticletitle{Auto-Sklearn 2.0: {H}ands-free AutoML via Meta-Learning}.
\newblock \bibinfo{journal}{\emph{Journal of Machine Learning Research}}  \bibinfo{volume}{23} (\bibinfo{year}{2022}), \bibinfo{pages}{261:1--261:61}.
\newblock


\bibitem[Feurer et~al\mbox{.}(2015)]%
        {FeuKleEggSpr+15}
\bibfield{author}{\bibinfo{person}{Matthias Feurer}, \bibinfo{person}{Aaron Klein}, \bibinfo{person}{Katharina Eggensperger}, \bibinfo{person}{Jost~Tobias Springenberg}, \bibinfo{person}{Manuel Blum}, {and} \bibinfo{person}{Frank Hutter}.} \bibinfo{year}{2015}\natexlab{}.
\newblock \showarticletitle{Efficient and Robust Automated Machine Learning}. In \bibinfo{booktitle}{\emph{Proc. of Advances in Neural Information Processing Systems (NeurIPS)}}.
\newblock


\bibitem[Flood et~al\mbox{.}(2024)]%
        {FloEngAspDes+24}
\bibfield{author}{\bibinfo{person}{Robert Flood}, \bibinfo{person}{Gints Engelen}, \bibinfo{person}{David Aspinall}, {and} \bibinfo{person}{Lieven Desmet}.} \bibinfo{year}{2024}\natexlab{}.
\newblock \showarticletitle{Bad Design Smells in Benchmark NIDS Datasets}. In \bibinfo{booktitle}{\emph{Proc. of the IEEE European Symposium on Security and Privacy (EuroS\&P)}}. \bibinfo{pages}{658--675}.
\newblock


\bibitem[Fortify(2013)]%
        {rats}
\bibfield{author}{\bibinfo{person}{Fortify}.} \bibinfo{year}{2013}\natexlab{}.
\newblock \bibinfo{title}{RATS on Google Code}.
\newblock \bibinfo{howpublished}{\url{https://code.google.com/archive/p/rough-auditing-tool-for-security/issues}}.
\newblock


\bibitem[Fu and Tantithamthavorn(2022)]%
        {FuCha+22}
\bibfield{author}{\bibinfo{person}{Michael Fu} {and} \bibinfo{person}{Chakkrit Tantithamthavorn}.} \bibinfo{year}{2022}\natexlab{}.
\newblock \showarticletitle{LineVul: A Transformer-based Line-Level Vulnerability Prediction}. In \bibinfo{booktitle}{\emph{Proceedings of the 19th International Conference on Mining Software Repositories (MSR)}}. \bibinfo{pages}{608--620}.
\newblock


\bibitem[Ganz et~al\mbox{.}(2023)]%
        {GanImgHärRie+23}
\bibfield{author}{\bibinfo{person}{Tom Ganz}, \bibinfo{person}{Erik Imgrund}, \bibinfo{person}{Martin Härterich}, {and} \bibinfo{person}{Konrad Rieck}.} \bibinfo{year}{2023}\natexlab{}.
\newblock \showarticletitle{PAVUDI: Patch-based Vulnerability Discovery using Machine Learning}. In \bibinfo{booktitle}{\emph{Proc. of the Annual Computer Security Applications Conference (ACSAC)}}. \bibinfo{pages}{704--717}.
\newblock


\bibitem[Gruska et~al\mbox{.}(2010)]%
        {GruWasZel10}
\bibfield{author}{\bibinfo{person}{Natalie Gruska}, \bibinfo{person}{Andrzej Wasylkowski}, {and} \bibinfo{person}{Andreas Zeller}.} \bibinfo{year}{2010}\natexlab{}.
\newblock \showarticletitle{Learning from 6,000 {P}rojects: {L}ightweight {C}ross-{P}roject {A}nomaly {D}etection}. In \bibinfo{booktitle}{\emph{Proc. of the International Symposium on Software Testing and Analysis (ISSTA)}}.
\newblock


\bibitem[Guo et~al\mbox{.}(2022)]%
        {GuoLuDuaWan+22}
\bibfield{author}{\bibinfo{person}{Daya Guo}, \bibinfo{person}{Shuai Lu}, \bibinfo{person}{Nan Duan}, \bibinfo{person}{Yanlin Wang}, \bibinfo{person}{Ming Zhou}, {and} \bibinfo{person}{Jian Yin}.} \bibinfo{year}{2022}\natexlab{}.
\newblock \showarticletitle{{UniXcoder}: {U}nified Cross-Modal Pre-training for Code Representation}. In \bibinfo{booktitle}{\emph{Proc. of the Annual Meeting of the Association for Computational Linguistics (ACL)}}.
\newblock


\bibitem[Halstead(1977)]%
        {Hal77}
\bibfield{author}{\bibinfo{person}{Maurice~Howard Halstead}.} \bibinfo{year}{1977}\natexlab{}.
\newblock \bibinfo{booktitle}{\emph{Elements of software science}}.
\newblock \bibinfo{publisher}{Elsevier}.
\newblock


\bibitem[Hamilton(2020)]%
        {Ham2020}
\bibfield{author}{\bibinfo{person}{William~L. Hamilton}.} \bibinfo{year}{2020}\natexlab{}.
\newblock \bibinfo{booktitle}{\emph{Graph {{Representation Learning}}}}.
\newblock \bibinfo{publisher}{Morgan \& Claypool Publishers}.
\newblock
\showISBNx{978-3-031-00460-5}


\bibitem[Harrison et~al\mbox{.}(1982)]%
        {HarMagKluc+82}
\bibfield{author}{\bibinfo{person}{Warren Harrison}, \bibinfo{person}{Kenneth Magel}, \bibinfo{person}{Raymond Kluczny}, {and} \bibinfo{person}{Arlan DeKock}.} \bibinfo{year}{1982}\natexlab{}.
\newblock \showarticletitle{Applying Software Complexity Metrics to Program Maintenance}.
\newblock \bibinfo{journal}{\emph{Computer}} \bibinfo{volume}{15}, \bibinfo{number}{9} (\bibinfo{year}{1982}), \bibinfo{pages}{65--79}.
\newblock


\bibitem[Hastie et~al\mbox{.}(2009)]%
        {HasTibFri09}
\bibfield{author}{\bibinfo{person}{Trevor Hastie}, \bibinfo{person}{Robert Tibshirani}, {and} \bibinfo{person}{Jerome Friedman}.} \bibinfo{year}{2009}\natexlab{}.
\newblock \bibinfo{booktitle}{\emph{The Elements of Statistical Learning: Data Mining, Inference and Prediction} (\bibinfo{edition}{second} ed.)}.
\newblock \bibinfo{publisher}{Springer}.
\newblock


\bibitem[Hill(1965)]%
        {Hil+65}
\bibfield{author}{\bibinfo{person}{Austin~Bradford Hill}.} \bibinfo{year}{1965}\natexlab{}.
\newblock \showarticletitle{The Environment and Disease: Association or Causation?}
\newblock \bibinfo{journal}{\emph{Proceedings of the Royal Society of Medicine}} (\bibinfo{year}{1965}).
\newblock


\bibitem[Imgrund et~al\mbox{.}(2023)]%
        {brokenPromises}
\bibfield{author}{\bibinfo{person}{Erik Imgrund}, \bibinfo{person}{Tom Ganz}, \bibinfo{person}{Martin H{\"{a}}rterich}, \bibinfo{person}{Lukas Pirch}, \bibinfo{person}{Niklas Risse}, {and} \bibinfo{person}{Konrad Rieck}.} \bibinfo{year}{2023}\natexlab{}.
\newblock \showarticletitle{Broken Promises: Measuring Confounding Effects in Learning-based Vulnerability Discovery}. In \bibinfo{booktitle}{\emph{Proc. of the {ACM} Workshop on Artificial Intelligence and Security (AISec)}}. \bibinfo{pages}{149--160}.
\newblock


\bibitem[Jimenez et~al\mbox{.}(2019)]%
        {JimRwePapSar+19}
\bibfield{author}{\bibinfo{person}{Matthieu Jimenez}, \bibinfo{person}{Renaud Rwemalika}, \bibinfo{person}{Mike Papadakis}, \bibinfo{person}{Federica Sarro}, \bibinfo{person}{Yves~Le Traon}, {and} \bibinfo{person}{Mark Harman}.} \bibinfo{year}{2019}\natexlab{}.
\newblock \showarticletitle{The importance of accounting for real-world labelling when predicting software vulnerabilities}. In \bibinfo{booktitle}{\emph{Proc. of the {ACM} Joint Meeting on European Software Engineering Conference (ESEC)}}.
\newblock


\bibitem[Joern Developer Community(2024)]%
        {joern}
Joern Developer Community \bibinfo{year}{2024}\natexlab{}.
\newblock \bibinfo{title}{Joern: The Bug Hunter's Workbench}.
\newblock
\urldef\tempurl%
\url{https://joern.io}
\showURL{%
\tempurl}


\bibitem[Juarez et~al\mbox{.}(2014)]%
        {JuaAfrAcaDía+14}
\bibfield{author}{\bibinfo{person}{Marc Juarez}, \bibinfo{person}{Sadia Afroz}, \bibinfo{person}{Gunes Acar}, \bibinfo{person}{Claudia Díaz}, {and} \bibinfo{person}{Rachel Greenstadt}.} \bibinfo{year}{2014}\natexlab{}.
\newblock \showarticletitle{A Critical Evaluation of Website Fingerprinting Attacks}. In \bibinfo{booktitle}{\emph{Proc. of the ACM Conference on Computer and Communications Security (CCS)}}. \bibinfo{pages}{263--274}.
\newblock


\bibitem[Larochelle and Evans(2001)]%
        {LarEva01}
\bibfield{author}{\bibinfo{person}{David Larochelle} {and} \bibinfo{person}{David Evans}.} \bibinfo{year}{2001}\natexlab{}.
\newblock \showarticletitle{Statically Detecting Likely Buffer Overflow Vulnerabilities}. In \bibinfo{booktitle}{\emph{Proc. of the USENIX Security Symposium}}.
\newblock


\bibitem[Liu et~al\mbox{.}(2020)]%
        {LiuMenZouGon+20}
\bibfield{author}{\bibinfo{person}{Bingchang Liu}, \bibinfo{person}{Guozhu Meng}, \bibinfo{person}{Wei Zou}, \bibinfo{person}{Qi Gong}, \bibinfo{person}{Feng Li}, \bibinfo{person}{Min Lin}, \bibinfo{person}{Dandan Sun}, \bibinfo{person}{Wei Huo}, {and} \bibinfo{person}{Chao Zhang}.} \bibinfo{year}{2020}\natexlab{}.
\newblock \showarticletitle{A large-scale empirical study on vulnerability distribution within projects and the lessons learned}. In \bibinfo{booktitle}{\emph{Proc. of the International Conference on Software Engineering (ICSE)}}. \bibinfo{pages}{1547--1559}.
\newblock


\bibitem[Liu et~al\mbox{.}(2024)]%
        {LiuTanZhaXia+24}
\bibfield{author}{\bibinfo{person}{Zhongxin Liu}, \bibinfo{person}{Zhijie Tang}, \bibinfo{person}{Junwei Zhang}, \bibinfo{person}{Xin Xia}, {and} \bibinfo{person}{Xiaohu Yang}.} \bibinfo{year}{2024}\natexlab{}.
\newblock \showarticletitle{Pre-training by Predicting Program Dependencies for Vulnerability Analysis Tasks}. In \bibinfo{booktitle}{\emph{Proc. of the International Conference on Software Engineering (ICSE)}}.
\newblock


\bibitem[Lozhkov et~al\mbox{.}(2024)]%
        {LoLi+24}
\bibfield{author}{\bibinfo{person}{Anton Lozhkov}, \bibinfo{person}{Raymond Li}, \bibinfo{person}{Loubna~Ben Allal}, \bibinfo{person}{Federico Cassano}, \bibinfo{person}{Joel Lamy{-}Poirier}, \bibinfo{person}{Nouamane Tazi}, \bibinfo{person}{Ao Tang}, \bibinfo{person}{Dmytro Pykhtar}, \bibinfo{person}{Jiawei Liu}, \bibinfo{person}{Yuxiang Wei}, \bibinfo{person}{Tianyang Liu}, \bibinfo{person}{Max Tian}, \bibinfo{person}{Denis Kocetkov}, \bibinfo{person}{Arthur Zucker}, \bibinfo{person}{Younes Belkada}, \bibinfo{person}{Zijian Wang}, \bibinfo{person}{Qian Liu}, \bibinfo{person}{Dmitry Abulkhanov}, \bibinfo{person}{Indraneil Paul}, \bibinfo{person}{Zhuang Li}, \bibinfo{person}{Wen{-}Ding Li}, \bibinfo{person}{Megan Risdal}, \bibinfo{person}{Jia Li}, \bibinfo{person}{Jian Zhu}, {et~al\mbox{.}}} \bibinfo{year}{2024}\natexlab{}.
\newblock \showarticletitle{StarCoder 2 and The Stack v2: The Next Generation}.
\newblock \bibinfo{journal}{\emph{Computing Research Repository (CoRR)}} (\bibinfo{year}{2024}).
\newblock


\bibitem[McCabe(1976)]%
        {McC76}
\bibfield{author}{\bibinfo{person}{Thomas~J. McCabe}.} \bibinfo{year}{1976}\natexlab{}.
\newblock \showarticletitle{A Complexity Measure}. In \bibinfo{booktitle}{\emph{Proc. of the International Conference on Software Engineering (ICSE)}}. \bibinfo{pages}{407}.
\newblock


\bibitem[Meng et~al\mbox{.}(2021)]%
        {MenGueMacAgh+21}
\bibfield{author}{\bibinfo{person}{Dongyu Meng}, \bibinfo{person}{Michele Guerriero}, \bibinfo{person}{Aravind Machiry}, \bibinfo{person}{Hojjat Aghakhani}, \bibinfo{person}{Priyanka Bose}, \bibinfo{person}{Andrea Continella}, \bibinfo{person}{Christopher Kruegel}, {and} \bibinfo{person}{Giovanni Vigna}.} \bibinfo{year}{2021}\natexlab{}.
\newblock \showarticletitle{Bran: Reduce Vulnerability Search Space in Large Open Source Repositories by Learning Bug Symptoms.}. In \bibinfo{booktitle}{\emph{Proc. of the ACM Asia Conference on Computer and Communications Security (AsiaCCS)}}. \bibinfo{pages}{731--743}.
\newblock


\bibitem[Nelder(1974)]%
        {loglinearRegression}
\bibfield{author}{\bibinfo{person}{J.~A. Nelder}.} \bibinfo{year}{1974}\natexlab{}.
\newblock \showarticletitle{Log Linear Models for Contingency Tables: A Generalization of Classical Least Squares}.
\newblock \bibinfo{journal}{\emph{Journal of the Royal Statistical Society. Series C (Applied Statistics)}} \bibinfo{volume}{23}, \bibinfo{number}{3} (\bibinfo{year}{1974}), \bibinfo{pages}{323--329}.
\newblock
\showISSN{00359254, 14679876}


\bibitem[Nijkamp et~al\mbox{.}(2023)]%
        {NiHaXiSa+23}
\bibfield{author}{\bibinfo{person}{Erik Nijkamp}, \bibinfo{person}{Hiroaki Hayashi}, \bibinfo{person}{Caiming Xiong}, \bibinfo{person}{Silvio Savarese}, {and} \bibinfo{person}{Yingbo Zhou}.} \bibinfo{year}{2023}\natexlab{}.
\newblock \showarticletitle{CodeGen2: Lessons for Training LLMs on Programming and Natural Languages}.
\newblock \bibinfo{journal}{\emph{Computing Research Repository (CoRR)}} (\bibinfo{year}{2023}).
\newblock


\bibitem[Nikitopoulos et~al\mbox{.}(2021)]%
        {NikDriLouMit21}
\bibfield{author}{\bibinfo{person}{Georgios Nikitopoulos}, \bibinfo{person}{Konstantina Dritsa}, \bibinfo{person}{Panos Louridas}, {and} \bibinfo{person}{Dimitris Mitropoulos}.} \bibinfo{year}{2021}\natexlab{}.
\newblock \showarticletitle{{CrossVul:} {A} Cross-Language Vulnerability Dataset with Commit Data}. In \bibinfo{booktitle}{\emph{Proc. of the {ACM} Joint European Software Engineering Conference~(ESEC)}}.
\newblock


\bibitem[of~Standards and (NIST)(2024)]%
        {sard}
\bibfield{author}{\bibinfo{person}{National~Institute of Standards} {and} \bibinfo{person}{Technology (NIST)}.} \bibinfo{year}{2024}\natexlab{}.
\newblock \bibinfo{title}{Software Assurance Reference Dataset (SARD)}.
\newblock
\urldef\tempurl%
\url{https://samate.nist.gov/SARD}
\showURL{%
\tempurl}


\bibitem[Park et~al\mbox{.}(2024)]%
        {PCV+24}
\bibfield{author}{\bibinfo{person}{Kiho Park}, \bibinfo{person}{Yo~Joong Choe}, {and} \bibinfo{person}{Victor Veitch}.} \bibinfo{year}{2024}\natexlab{}.
\newblock \showarticletitle{The Linear Representation Hypothesis and the Geometry of Large Language Models}. In \bibinfo{booktitle}{\emph{Proc. of the International Conference on Machine Learning (ICML)}}.
\newblock


\bibitem[Pearl(2009)]%
        {Pea09}
\bibfield{author}{\bibinfo{person}{Judea Pearl}.} \bibinfo{year}{2009}\natexlab{}.
\newblock \bibinfo{booktitle}{\emph{Causality: Models, Reasoning and Inference} (\bibinfo{edition}{2nd} ed.)}.
\newblock \bibinfo{publisher}{Cambridge University Press}.
\newblock


\bibitem[Perl et~al\mbox{.}(2015)]%
        {PerDecSmiArp+15}
\bibfield{author}{\bibinfo{person}{Henning Perl}, \bibinfo{person}{Sergej Dechand}, \bibinfo{person}{Matthew Smith}, \bibinfo{person}{Daniel Arp}, \bibinfo{person}{Fabian Yamaguchi}, \bibinfo{person}{Konrad Rieck}, \bibinfo{person}{Sascha Fahl}, {and} \bibinfo{person}{Yasemin Acar}.} \bibinfo{year}{2015}\natexlab{}.
\newblock \showarticletitle{VCCFinder: Finding Potential Vulnerabilities in Open-Source Projects to Assist Code Audits}. In \bibinfo{booktitle}{\emph{Proc. of the ACM Conference on Computer and Communications Security (CCS)}}. \bibinfo{pages}{426--437}.
\newblock


\bibitem[Platonov et~al\mbox{.}(2023)]%
        {PlaKuzDisBab+23}
\bibfield{author}{\bibinfo{person}{Oleg Platonov}, \bibinfo{person}{Denis Kuznedelev}, \bibinfo{person}{Michael Diskin}, \bibinfo{person}{Artem Babenko}, {and} \bibinfo{person}{Liudmila Prokhorenkova}.} \bibinfo{year}{2023}\natexlab{}.
\newblock \showarticletitle{A critical look at the evaluation of GNNs under heterophily: Are we really making progress?}. In \bibinfo{booktitle}{\emph{Proc. of the International Conference on Learning Representations (ICLR)}}.
\newblock


\bibitem[Rice(1953)]%
        {Ric53}
\bibfield{author}{\bibinfo{person}{Henry~Gordon Rice}.} \bibinfo{year}{1953}\natexlab{}.
\newblock \showarticletitle{Classes of recursively enumerable sets and their decision problems}.
\newblock \bibinfo{journal}{\emph{Trans. Amer. Math. Soc.}}  \bibinfo{volume}{74} (\bibinfo{year}{1953}), \bibinfo{pages}{358--366}.
\newblock


\bibitem[Risse and Böhme(2024)]%
        {RisBöh24}
\bibfield{author}{\bibinfo{person}{Niklas Risse} {and} \bibinfo{person}{Marcel Böhme}.} \bibinfo{year}{2024}\natexlab{}.
\newblock \showarticletitle{Uncovering the Limits of Machine Learning for Automatic Vulnerability Detection}. In \bibinfo{booktitle}{\emph{Proc. of the USENIX Security Symposium}}.
\newblock


\bibitem[Risse et~al\mbox{.}(2025)]%
        {RisBoe24}
\bibfield{author}{\bibinfo{person}{Niklas Risse}, \bibinfo{person}{Jing Liu}, {and} \bibinfo{person}{Marcel B\"{o}hme}.} \bibinfo{year}{2025}\natexlab{}.
\newblock \showarticletitle{Top Score on the Wrong Exam: On Benchmarking in Machine Learning for Vulnerability Detection}.
\newblock \bibinfo{journal}{\emph{Proc. {ACM} Softw. Eng.}} \bibinfo{volume}{2}, \bibinfo{number}{ISSTA} (\bibinfo{year}{2025}), \bibinfo{pages}{388--410}.
\newblock


\bibitem[Russell et~al\mbox{.}(2018)]%
        {RusKimHamLaz+18}
\bibfield{author}{\bibinfo{person}{Rebecca~L. Russell}, \bibinfo{person}{Louis~Y. Kim}, \bibinfo{person}{Lei~H. Hamilton}, \bibinfo{person}{Tomo Lazovich}, \bibinfo{person}{Jacob Harer}, \bibinfo{person}{Onur Ozdemir}, \bibinfo{person}{Paul~M. Ellingwood}, {and} \bibinfo{person}{Marc~W. McConley}.} \bibinfo{year}{2018}\natexlab{}.
\newblock \showarticletitle{Automated Vulnerability Detection in Source Code Using Deep Representation Learning}. In \bibinfo{booktitle}{\emph{Proc. of the {IEEE} International Conference on Machine Learning and Applications~(ICMLA)}}.
\newblock


\bibitem[Scandariato et~al\mbox{.}(2014)]%
        {ScaWlHovJoo14}
\bibfield{author}{\bibinfo{person}{Riccardo Scandariato}, \bibinfo{person}{James Walden}, \bibinfo{person}{Aram Hovsepyan}, {and} \bibinfo{person}{Wouter Joosen}.} \bibinfo{year}{2014}\natexlab{}.
\newblock \showarticletitle{Predicting Vulnerable Software Components via Text Mining}.
\newblock \bibinfo{journal}{\emph{IEEE Transactions on Software Engineering}} \bibinfo{volume}{40}, \bibinfo{number}{10} (\bibinfo{year}{2014}), \bibinfo{pages}{993--1006}.
\newblock


\bibitem[Sejfia et~al\mbox{.}(2024)]%
        {SejDasShaMed24}
\bibfield{author}{\bibinfo{person}{Adriana Sejfia}, \bibinfo{person}{Satyaki Das}, \bibinfo{person}{Saad Shafiq}, {and} \bibinfo{person}{Nenad Medvidovic}.} \bibinfo{year}{2024}\natexlab{}.
\newblock \showarticletitle{Toward Improved Deep Learning-based Vulnerability Detection}. In \bibinfo{booktitle}{\emph{Proc. of the International Conference on Software Engineering (ICSE)}}.
\newblock


\bibitem[Semgrep, Inc.(2025)]%
        {semgrep}
Semgrep, Inc. \bibinfo{year}{2025}\natexlab{}.
\newblock \bibinfo{booktitle}{\emph{Semgrep}}.
\newblock
\urldef\tempurl%
\url{https://semgrep.dev}
\showURL{%
\tempurl}


\bibitem[Shin and Williams(2008)]%
        {ShiWil08}
\bibfield{author}{\bibinfo{person}{Yonghee Shin} {and} \bibinfo{person}{Laurie Williams}.} \bibinfo{year}{2008}\natexlab{}.
\newblock \showarticletitle{An Empirical Model to Predict Security Vulnerabilities using Code Complexity Metrics}. In \bibinfo{booktitle}{\emph{Proc. of the International Symposium on Empirical Software Engineering and Measurement}}. \bibinfo{pages}{315–317}.
\newblock


\bibitem[Shin and Williams(2013)]%
        {ShiWil+13}
\bibfield{author}{\bibinfo{person}{Yonghee Shin} {and} \bibinfo{person}{Laurie Williams}.} \bibinfo{year}{2013}\natexlab{}.
\newblock \showarticletitle{Can traditional fault prediction models be used for vulnerability prediction?}
\newblock \bibinfo{journal}{\emph{Empirical Software Engineering}} (\bibinfo{year}{2013}).
\newblock
\showISSN{1573-7616}


\bibitem[Sommer and Paxson(2010)]%
        {SomPax10}
\bibfield{author}{\bibinfo{person}{Robin Sommer} {and} \bibinfo{person}{Vern Paxson}.} \bibinfo{year}{2010}\natexlab{}.
\newblock \showarticletitle{Outside the Closed World: On Using Machine Learning for Network Intrusion Detection}. In \bibinfo{booktitle}{\emph{Proc. of the IEEE Symposium on Security and Privacy (S\&P)}}. \bibinfo{pages}{305--316}.
\newblock


\bibitem[Tree-sitter(2024)]%
        {treesitter}
Tree-sitter \bibinfo{year}{2024}\natexlab{}.
\newblock \bibinfo{booktitle}{\emph{Tree-Sitter: Parser Generator Tool}}.
\newblock
\urldef\tempurl%
\url{https://tree-sitter.github.io/tree-sitter/}
\showURL{%
\tempurl}


\bibitem[Ullah et~al\mbox{.}(2024)]%
        {UllHanPujPea+24}
\bibfield{author}{\bibinfo{person}{Saad Ullah}, \bibinfo{person}{Mingji Han}, \bibinfo{person}{Saurabh Pujar}, \bibinfo{person}{Hammond Pearce}, \bibinfo{person}{Ayse~K. Coskun}, {and} \bibinfo{person}{Gianluca Stringhini}.} \bibinfo{year}{2024}\natexlab{}.
\newblock \showarticletitle{LLMs Cannot Reliably Identify and Reason About Security Vulnerabilities (Yet?): A Comprehensive Evaluation, Framework, and Benchmarks.}. In \bibinfo{booktitle}{\emph{Proc. of the IEEE Symposium on Security and Privacy (S\&P)}}. \bibinfo{pages}{862--880}.
\newblock


\bibitem[Wagner et~al\mbox{.}(2000)]%
        {WagFosBreAik+00}
\bibfield{author}{\bibinfo{person}{David~A. Wagner}, \bibinfo{person}{Jeffrey~S. Foster}, \bibinfo{person}{Eric~A. Brewer}, {and} \bibinfo{person}{Alexander Aiken}.} \bibinfo{year}{2000}\natexlab{}.
\newblock \showarticletitle{A First Step Towards Automated Detection of Buffer Overrun Vulnerabilities}. In \bibinfo{booktitle}{\emph{Proc. of the Network and Distributed System Security Symposium (NDSS)}}.
\newblock


\bibitem[Wang et~al\mbox{.}(2023)]%
        {WanLeGotBui+23}
\bibfield{author}{\bibinfo{person}{Yue Wang}, \bibinfo{person}{Hung Le}, \bibinfo{person}{Akhilesh Gotmare}, \bibinfo{person}{Nghi D.~Q. Bui}, \bibinfo{person}{Junnan Li}, {and} \bibinfo{person}{Steven C.~H. Hoi}.} \bibinfo{year}{2023}\natexlab{}.
\newblock \showarticletitle{{CodeT5+:} Open Code Large Language Models for Code Understanding and Generation}. In \bibinfo{booktitle}{\emph{Proc. of the Conference on Empirical Methods in Natural Language Processing (EMNLP)}}.
\newblock


\bibitem[Wang et~al\mbox{.}(2021)]%
        {WanWanJotHoi21}
\bibfield{author}{\bibinfo{person}{Yue Wang}, \bibinfo{person}{Weishi Wang}, \bibinfo{person}{Shafiq~R. Joty}, {and} \bibinfo{person}{Steven C.~H. Hoi}.} \bibinfo{year}{2021}\natexlab{}.
\newblock \showarticletitle{{CodeT5:} {I}dentifier-aware Unified Pre-trained Encoder-Decoder Models for Code Understanding and Generation}. In \bibinfo{booktitle}{\emph{Proc. of the Conference on Empirical Methods in Natural Language Processing (EMNLP)}}.
\newblock


\bibitem[Weissberg et~al\mbox{.}(2024)]%
        {WeiMölGanImg+24}
\bibfield{author}{\bibinfo{person}{Felix Weissberg}, \bibinfo{person}{Jonas Möller}, \bibinfo{person}{Tom Ganz}, \bibinfo{person}{Erik Imgrund}, \bibinfo{person}{Lukas Pirch}, \bibinfo{person}{Lukas Seidel}, \bibinfo{person}{Moritz Schloegel}, \bibinfo{person}{Thorsten Eisenhofer}, {and} \bibinfo{person}{Konrad Rieck}.} \bibinfo{year}{2024}\natexlab{}.
\newblock \showarticletitle{SoK: Where to Fuzz? Assessing Target Selection Methods in Directed Fuzzing}. In \bibinfo{booktitle}{\emph{Proc. of the ACM Asia Conference on Computer and Communications Security (AsiaCCS)}}.
\newblock


\bibitem[Wressnegger et~al\mbox{.}(2016)]%
        {WreYamMaiRie+16}
\bibfield{author}{\bibinfo{person}{Christian Wressnegger}, \bibinfo{person}{Fabian Yamaguchi}, \bibinfo{person}{Alwin Maier}, {and} \bibinfo{person}{Konrad Rieck}.} \bibinfo{year}{2016}\natexlab{}.
\newblock \showarticletitle{Twice the Bits, Twice the Trouble: Vulnerabilities Induced by Migrating to 64-Bit Platforms}. In \bibinfo{booktitle}{\emph{Proc. of the ACM Conference on Computer and Communications Security (CCS)}}. \bibinfo{pages}{541--552}.
\newblock


\bibitem[Wright(1921)]%
        {Wri+21}
\bibfield{author}{\bibinfo{person}{Sewall Wright}.} \bibinfo{year}{1921}\natexlab{}.
\newblock \showarticletitle{Correlation and causation}.
\newblock \bibinfo{journal}{\emph{Journal of agricultural research}} \bibinfo{volume}{20}, \bibinfo{number}{7} (\bibinfo{year}{1921}), \bibinfo{pages}{557}.
\newblock


\bibitem[Yamaguchi et~al\mbox{.}(2014)]%
        {YamGolArpRie+14}
\bibfield{author}{\bibinfo{person}{Fabian Yamaguchi}, \bibinfo{person}{Nico Golde}, \bibinfo{person}{Daniel Arp}, {and} \bibinfo{person}{Konrad Rieck}.} \bibinfo{year}{2014}\natexlab{}.
\newblock \showarticletitle{Modeling and Discovering Vulnerabilities with Code Property Graphs}. In \bibinfo{booktitle}{\emph{Proc. of the IEEE Symposium on Security and Privacy (S\&P)}}. \bibinfo{pages}{590--604}.
\newblock


\bibitem[Yamaguchi et~al\mbox{.}(2012)]%
        {YamLotRie12}
\bibfield{author}{\bibinfo{person}{Fabian Yamaguchi}, \bibinfo{person}{Markus Lottmann}, {and} \bibinfo{person}{Konrad Rieck}.} \bibinfo{year}{2012}\natexlab{}.
\newblock \showarticletitle{Generalized vulnerability extrapolation using abstract syntax trees}. In \bibinfo{booktitle}{\emph{Proc. of the Annual Computer Security Applications Conference (ACSAC)}}. \bibinfo{pages}{359--368}.
\newblock


\bibitem[Zhou et~al\mbox{.}(2019)]%
        {ZhoLiuSioDu+19}
\bibfield{author}{\bibinfo{person}{Yaqin Zhou}, \bibinfo{person}{Shangqing Liu}, \bibinfo{person}{Jing~Kai Siow}, \bibinfo{person}{Xiaoning Du}, {and} \bibinfo{person}{Yang Liu}.} \bibinfo{year}{2019}\natexlab{}.
\newblock \showarticletitle{Devign: Effective Vulnerability Identification by Learning Comprehensive Program Semantics via Graph Neural Networks}. In \bibinfo{booktitle}{\emph{Proc. of the Conference on Neural Information Processing Systems (NeurIPS)}}.
\newblock


\end{thebibliography}

\clearpage
\appendix
\section{Structural Code Metrics}
\label{sec:appendix-metrics}

In our study, we analyze and implement 23 code metrics within the framework introduced in \cref{sec:codemetrics}. For each metric, we provide a formal definition of the filter function $\filter$, mapping function $\map$, and reduction function $\reduce$ in \cref{tab:cm-examples}. We use $\mathbbm{1}$ as the indicator function. Some metrics use auxiliary measures, which are denoted with an additional index, such as M1.1.

For the filtering step, we traverse the syntax tree to identify and extract all subtrees that match specific code metric patterns. These patterns are defined using a tree query language capturing node types, their relationships, and apply quantifiers to these relationships. We build on the Tree-sitter query language~\cite{treesitter}, extending it to provide the necessary expressiveness.
Conceptually, our language is based on \emph{S-expressions}, which are textual representations of tree structures originating from the LISP programming language. Just as regular expressions used for pattern matching in text, this query language enables pattern matching within tree structures. \cref{tab:lang-def} lists the building blocks of our query language. 

The core unit of our language is a node, which can be matched by a specific type, a wildcard specifier \sexpinline{(\_)}, an alternative of types \sexpinline{((a) | (b))}, or a type negation \sexpinline{(!a)}. 
Each node can be connected by three types of relationships: parent-child, siblings, and descendants. The parent-child relationship is specified by placing the child node within the parentheses of the parent, as in \sexpinline{(a (b))}.

\begin{table*}[b!]
    \centering
    \caption{Overview over all 23 structural code metrics.\vspace{-0.25cm}}
    \centering
    \scalebox{0.93}{
    \footnotesize
\begin{tabular}{@{} p{0.025\textwidth} p{0.248\textwidth} p{0.38\textwidth} p{0.16\textwidth} c @{}}
\toprule
 & \multicolumn{1}{c}{\textbf{Description}} & \multicolumn{1}{c}{\textbf{Filter}} & \multicolumn{1}{c}{\textbf{Map}} & \multicolumn{1}{c}{\phantom{  }\textbf{Reduce}} \\
\midrule
\multicolumn{5}{c}{\textbf{Code Smell Metrics}} \\
S1
    & \# magic numbers
    & \begin{sexp}
(number_literal) @num
\end{sexp}
    & $\subtree \mapsto \mathbbm{1} [ \subtree.\text{num} \notin \{-1,0,1\} ]$
    & sum \\
S2
    & \# goto
    & \begin{sexp}
(goto_stmt)
\end{sexp}
    & $\subtree \mapsto 1$
    & sum \\
S3
    & \# function pointers
    & \begin{sexp}
((declaration (init_declarator (function_declarator)))
| (parameter_declaration (function_declarator)))
\end{sexp}
    & $\subtree \mapsto 1$
    & sum \\
S4
    & \# function calls with unused returns
    & \begin{sexp}
(expr_stmt (call_expr))
\end{sexp}
    & $\subtree \mapsto 1$
    & sum \\
S5
    & \# if without else
    & \begin{sexp}
(if_stmt !alternative)
\end{sexp}
    & $\subtree \mapsto 1$
    & sum \\
\midrule
\multicolumn{5}{c}{\textbf{Complexity Metrics}} \\
C1
    & cyclomatic complexity
    & \begin{sexp}
(cond_stmt (_))
\end{sexp}
    & $\subtree\mapsto 1 + \text{c1.1}(\subtree)$
    & sum \\
C1.1
    & \# logical operators
    & \begin{sexp}
(binary_expr (operator) @op)
\end{sexp}
    & $\subtree\mapsto \mathbbm{1} [ \subtree.\text{op} \in \{\text{'\&\&'}, \text{'}||\text{'}\} ]$
    & sum \\
C2
    & \# loops
    & \begin{sexp}
(loop_stmt)
\end{sexp}
    & $\subtree \mapsto 1$
    & sum \\
C3
    & \# nested loops
    & \begin{sexp}
(loop_stmt ((!loop_stmt)$\hat{*}$ (loop_stmt)))
\end{sexp}
    & $\subtree \mapsto 1$
    & sum \\
C4
    & max nesting level of loops
    & \begin{sexp}
(loop_stmt ((!loop_stmt)$\hat{*}$ (loop_stmt))$\hat{*}$)
\end{sexp}
    & $\subtree \mapsto \text{C2}(\subtree) + 1$
    & max \\
C5
    & \# parameters
    & \begin{sexp}
(parameter_declaration)
\end{sexp}
    & $\subtree \mapsto 1$
    & max \\
C6
    & \# nested control structures
    & \begin{sexp}
(ctrl_stmt ((!ctrl_stmt)$\hat{*}$ (ctrl_stmt)))
\end{sexp}
    & $\subtree \mapsto 1$
    & sum \\
C7
    & max nesting level of control structure
    & \begin{sexp}
(ctrl_stmt ((!ctrl_stmt)$\hat{*}$ (ctrl_stmt))$\hat{*}$)
\end{sexp}
    & $\subtree \mapsto \text{C6}(\subtree) + 1$
    & max \\
C8
    & max \# control structures in a control structure
    & \begin{sexp}
(ctrl_stmt ((!ctrl_stmt)$_{*}^{\hat{*}}$ (ctrl_stmt)$_{*}^{\hat{*}}$)
\end{sexp}
    & $\subtree \mapsto \text{C6}(\subtree)$
    & max \\
C9
    & \# return stmts
    & \begin{sexp}
(return_stmt)
\end{sexp}
    & $\subtree \mapsto 1$
    & sum \\
C10
    & \# casts
    & \begin{sexp}
(cast_expr)
\end{sexp}
    & $\subtree \mapsto 1$
    & sum \\
C11
    & \# local variables
    & \begin{sexp}
(declaration)
\end{sexp}
    & $\subtree \mapsto 1$
    & sum \\
C12
    & max \# operands
    & \begin{sexp}
(binary_expr)
\end{sexp}
    & $\subtree \mapsto \mu_{\text{(identifier$\mid$literal)},\text{sum}}(\subtree)$
    & max \\
\midrule
\multicolumn{5}{c}{\textbf{Memory Metrics}} \\
M1
    & \# heap allocations
    & \begin{sexp}
(_)
\end{sexp}
    & $\subtree \mapsto \text{M}1.1(\subtree) + \text{M}1.2(\subtree)$
    & sum \\
M1.1
    & \# new allocations
    & \begin{sexp}
(new_expr)
\end{sexp}
    & $\subtree \mapsto 1$
    & sum \\
M1.2
    & \# call allocations
    & \begin{sexp}
(call_expr function: (identifier) @name)
\end{sexp}
    & $\subtree\mapsto \mathbbm{1} [ \text{'alloc' in \,} \subtree.\text{name} ]$
    & sum \\
M2
    & \# pointer dereferences
    & \begin{sexp}
((pointer_expr) | (subscript_expr) | (field_expr))
\end{sexp}
    & $\subtree \mapsto 1$
    & sum \\
M3 
    & \# pointer arithmetic
    & \begin{sexp}
((binary_expr) | (unary_expr))
\end{sexp}
    & $\subtree \mapsto 
\mathbbm{1} [ \text{M3.1}(\subtree) > 0 ]
$
    & sum \\
M3.1
    & \# pointer variables
    & \begin{sexp}
({type: 'pointer'})
\end{sexp}
    & $\subtree \mapsto 1$
    & sum \\
\midrule
\multicolumn{5}{c}{\textbf{Syntax Tree Metrics}} \\
T1
    & \# AST nodes
    & \begin{sexp}
(_)
\end{sexp}
    & $\subtree \mapsto 1$
    & sum \\
T2
    & max height of AST
    & \begin{sexp}
(_ (_)$\hat{*}$)
\end{sexp}
    & $\subtree \mapsto \lvert \subtree \rvert$
    & max \\
T3
    & average \# of children
    & \begin{sexp}
(_ (_)$+$ @children)
\end{sexp}
    & $\subtree \mapsto \lvert \subtree.\text{children} \rvert$
    & avg \\
 \bottomrule
\end{tabular}

    }
    \label{tab:cm-examples}
\end{table*}

\begin{table}[H]
    \centering
    \footnotesize
    \caption{Building structures for our tree query language.\vspace{-0.25cm}}
    \label{tab:lang-def}
    \scalebox{0.96}{
    \begin{tabular}{@{}llp{0.53\columnwidth}@{}}
    \toprule
    \textbf{Rule}                  & \textbf{Notation}              & \textbf{Description} \\
    \midrule
    Node                  & \sexpinline{(a)}              & Matches each node of type `a' \\
    Parent-Child          & \sexpinline{(a (b))}          & \dots with one child of type `b' \\
    Siblings       & \sexpinline{(a (b) (c))}      & \dots with exactly two children of type `b' and `c'\\
    Sibling Quantifier    & \sexpinline{(a (b)*)}         & \dots with arbitrarily many children of type `b'\\
    Descendant Quantifier & \sexpinline{(a (b)$\hat{*}$)} & \dots with a path graph of nodes of type `b'\\
    Alternative           & \sexpinline{((a) | (b))}      & Matches each node of type `a' or `b' \\
    Negation              & \sexpinline{(\!a)}             & Matches each node whose type is not `a' \\
    Wildcard              & \sexpinline{(\_)}             & Matches any node \\
    Annotation            & \sexpinline{(a) @x}        & Assigns the name 'x' to node of type `a' \\
    \bottomrule
    \end{tabular}}
\end{table}

Adding more child nodes implies a sibling relationship, as in \sexpinline{(a (b) (c))}, where \sexpinline{b} and \sexpinline{c} are siblings. To indicate an arbitrary number of children, we use sibling quantifiers \sexpinline{*} (zero or more matches) and \sexpinline{+} (one or more matches) on a child node. Similarly, the descendant quantifiers \sexpinline{$\hat{*}$} and \sexpinline{$\hat{+}$} allow expressing depth-traversal of the tree.
Lastly, to access attributes of specific nodes in the mapping function, they can be annotated with \sexpinline{(a) @node}. The following map function can use this annotation to calculate various numerical values. For example, $\vert \subtree.\text{node} \vert$ corresponds to the cardinality of the node type in the subtree~$\subtree$.

\clearpage

\section{Additional Results}
\cbstart

In addition to the main results presented in the paper, we aim to provide further insight into the model configurations and a more fine-grained view of our results. 

\paragraph{Model configuration}
The transformer-based models were trained for 10 epochs using a learning rate of $2 \times 10^{-5}$, a batch size of 8 with 8 gradient accumulation steps, and early stopping with a patience of 10 and no minimal improvement requirement. ReVeal and Devign used larger batch sizes of 256 with 50 gradient accumulation steps to stabilize optimization, a higher learning rate of $1 \times 10^{-4}$ and were trained for 100 epochs with the same early stopping criterion. The code metrics-based model was trained for 10 minutes, requiring no further hyper parameter selection.

\begin{table*}[t]
    \centering\small
    \caption{\cb{Performance of different vulnerability prediction approaches on the PrimeVul~\citep{DinFuIbrSit+25} dataset.}}
    \cbstart
\setlength{\tabcolsep}{6pt}
\begin{tabular}{l r r r r r r }
    \toprule
    \textbf{Predictor} & \textbf{Efficiency} & \textbf{F1 $\uparrow$} & \textbf{AUPRC $\uparrow$} & \textbf{MCC $\uparrow$} & \textbf{BAcc $\uparrow$} & \textbf{VD-S $\downarrow$} \\
    \midrule

    \multicolumn{7}{l}{\emph{Code Metrics}} \\
     \phantom{xx}\scmAuto{} & 2.354 & \secondbest{20.32} $\pm$ 0.59 & \best{13.80} $\pm$ 0.82 & 18.90 $\pm$ 0.55 & 62.22 $\pm$ 0.64 & \best{90.20} $\pm$ 0.70 \\

    \midrule
    
    \multicolumn{7}{l}{\emph{Code Language Models}} \\
    \rule{0pt}{2ex}
    \phantom{xx}UniXcoder~\cite{GuoLuDuaWan+22} & 0.130 & \best{20.69} $\pm$ 1.43 & 13.32 $\pm$ 1.22 & \best{21.36} $\pm$ 1.06 & 67.23 $\pm$ 2.38 & 92.84 $\pm$ 0.79 \\
    \phantom{xx}PDBERT~\cite{LiuTanZhaXia+24} & 0.121 & 19.50 $\pm$ 1.17 & \secondbest{13.38} $\pm$ 0.68 & \secondbest{20.67} $\pm$ 0.40 & \secondbest{69.19} $\pm$ 1.86 & \secondbest{92.40} $\pm$ 1.23 \\
    \phantom{xx}CodeGen 2.5~\cite{NiHaXiSa+23} & 0.002 & 18.57 $\pm$ 0.54 & 13.22 $\pm$ 0.30 & 17.07 $\pm$ 0.41 & 63.23 $\pm$ 1.19 & 92.50 $\pm$ 0.49 \\
    \phantom{xx}LineVul~\cite{FuCha+22} & 0.112 & 18.48 $\pm$ 1.23 & 11.68 $\pm$ 0.82 & 19.34 $\pm$ 0.93 & 68.36 $\pm$ 2.06 & 92.66 $\pm$ 1.35 \\
    \phantom{xx}StarCoder 2~\cite{LoLi+24} & 0.002 & 17.09 $\pm$ 0.41 & 12.46 $\pm$ 0.40 & 19.73 $\pm$ 0.48 & \best{72.25} $\pm$ 1.41 & 94.17 $\pm$ 0.44 \\

    \midrule
    
    \multicolumn{7}{l}{\emph{General-purpose LMs}} \\ 
    \phantom{xx}GPT-3.5 Turbo & $1.3 \times 10^{-4}$ & 5.80 $\pm$ 0.83 & 2.56 $\pm$ 0.18 & 3.34 $\pm$ 1.01 & 52.96 $\pm$ 1.09 & 100.00 $\pm$ 0.00 \\
    \phantom{xx}GPT-4o & $8.90 \times 10^{-7}$ & 5.31 $\pm$ 0.35 & 2.50 $\pm$ 0.10 & 2.72 $\pm$ 0.90 & 54.17 $\pm$ 1.47 & 100.00 $\pm$ 0.00 \\
    \phantom{xx}Random & \multicolumn{1}{c}{-} & 4.42 $\pm$ 0.19 & 2.33 $\pm$ 0.01 &  0.05 $\pm$ 0.36 & 50.15 $\pm$ 0.45 & 99.60 $\pm$ 0.18 \\

    \bottomrule
\end{tabular}
\setlength{\tabcolsep}{6pt}
\cbend
    \label{tab:class_results_extended}
\end{table*}

\paragraph{Model performance}
In addition to the performance metrics reported as part of our evaluation in \cref{tab:class_results}, we further report in \cref{tab:class_results_extended} the Matthews correlation coefficient (MCC), the balanced accuracy (BAcc), and the vulnerability detection score (VD-S)---the false negative rate at a fixed false positive rate of 0.05\%---introduced alongside the PrimeVul dataset~\citep{DinFuIbrSit+25}.

Most importantly, we find that no single model achieves the best performance across all metrics. However, three models stand out: UniXcoder ranks highest in two of the five metrics but never ranks second; PDBERT ranks second-best in four of the five metrics but never ranks highest; and the code metrics-based model achieves the best score in two metrics and the second-best score in one.

This finding suggests that analyses of vulnerability discovery models should always report results across a variety of metrics, as the assessment of which model performs best can vary depending on the chosen metric. In our experiments, UniXcoder, PDBERT, and the code metrics-based model perform comparably overall considering different metrics.

In addition to the model performance, we report the parameter efficiency by calculating the improvement in percentage points over the random baseline per million parameters. This metric addresses the intuitive question: How much improvement over the baseline is gained by adding one million parameters? We find that the code metrics-based model shows the greatest parameter efficiency at two orders of magnitude higher than those of the next best, UniXCoder, which shows the greatest efficiency among the LLM-based approaches. This metric can guide the selection of models that offer the best trade-off between performance and size.

\clearpage
\cbend

\end{document}